\documentclass[10pt,journal,compsoc]{IEEEtran}
\ifCLASSOPTIONcompsoc
  \usepackage[nocompress]{cite}
\else
  \usepackage{cite}
\fi
\ifCLASSINFOpdf
  \usepackage[pdftex]{graphicx}
\else
  \usepackage[dvips]{graphicx}
\fi
\usepackage{amsmath}
\usepackage{url}

\usepackage{multirow}
\usepackage{color,soul}
\usepackage{framed} 

\begin{document}
%
\title{Systematic Evaluation and Usability Analysis of Formal Tools for Railway System Design}
%
%
%
%

\author{Alessio~Ferrari, Franco~Mazzanti, Davide~Basile, and~Maurice~H.~ter~Beek
\IEEEcompsocitemizethanks{\IEEEcompsocthanksitem The authors are with the Institute of Information Science and Technologies (ISTI) of the Italian National Research Council (CNR), Pisa, Italy. 
E-mail: \{alessio.ferrari,franco.mazzanti,davide.basile,maurice.terbeek\}@isti.cnr.it}
\thanks{Manuscript received April, 2021; revised \ldots}}

%
%

\markboth{Submitted to IEEE Transactions on Software Engineering, April~2021}%
{Ferrari \MakeLowercase{\textit{et al.}}: Systematic Evaluation and Usability Analysis of Formal Tools for Railway System Design}
%



\IEEEtitleabstractindextext{%
\begin{abstract}
Formal methods and supporting tools have a long record of successes in the development of safety-critical systems. However, no single tool has emerged as the dominant solution for system design. Each tool differs from the others in terms of the modeling language used, its verification capabilities and other complementary features, and each development context has peculiar needs that require different tools. This is particularly problematic for the railway industry, in which formal methods are highly recommended by the norms, but no actual guidance is provided for the selection of tools. To guide companies in the selection of the most appropriate formal tools to adopt in their contexts, a clear assessment of the features of the currently available tools is required. 

To address this goal, this paper considers a set of 13~formal tools that have been used for railway system design, and it presents a systematic evaluation of such tools and a preliminary usability analysis of a subset of 7~tools, involving railway practitioners. The results are discussed considering the most desired aspects by industry and earlier related studies. While the focus is on the railway domain, the overall methodology can be applied to similar contexts. Our study thus contributes with a systematic evaluation of formal tools and it shows that despite the poor graphical interfaces, \textit{usability} and \textit{maturity} of the tools are not major problems, as claimed by contributions from the literature. Instead, support for \textit{process integration} is the most relevant obstacle for adoption of most of the tools.
\end{abstract}

}

\maketitle


%

\IEEEraisesectionheading{\section{Introduction}\label{sec:introduction}}

%
%
%
%
\IEEEPARstart{T}{he} development of railway safety-critical systems, such as platforms for on-board automatic train control~\cite{ChiappiniCMRRSTV10,ferrari2013metro} or computer-based interlocking infrastructures to route the trains~\cite{HaxthausenPK11,WR03}, has to follow strict process guidelines to deliver products that are highly dependable and trustworthy~\cite{fantechi2013twenty,FFG16}. Formal methods are mathematics-based techniques for the specification, development and (manual or automated) verification of software and hardware systems~\cite{wing1990specifier,GBP20}, and are particularly indicated when rigor is a main concern. They have a long history of over 30 years of success stories in railway applications~\cite{fantechi2013twenty,woodcock2009formal,garavel2013}, with several support tools available in the market~\cite{garavel2019reflections,FBMBFGPT19,hartmanns2015quantitative,katoen2016probabilistic}.   
Furthermore, the CENELEC EN~50128 norm~\cite{cenelec50128}, which is the standard for the development of railway software in Europe, 
highly recommends the usage of formal methods for the design and verification of those products that need to meet the highest safety integrity levels. 

Despite these premises, the adoption of formal methods and their supporting tools by companies is rather limited~\cite{davis2013study,nyberg2018formal}, and railway practitioners ask for more guidance to select the most adequate formal tool, or set of tools, for their development contexts~\cite{FMBBF20,schlick2018proposal,steffen2017physics}. This is common also to other application domains. As observed by Steffen~\cite{steffen2017physics}:  ``Prospective users have a hard time to orient themselves in the current tool landscape, and even experts typically only have very partial knowledge. Thus, the need for a more systematic approach to establish the profiles of tools and methods is obvious''.  

Previous work on applications of formal methods to railway problems has mostly focused on reporting experiences~\cite{WR03,abrial2007formal,ChiappiniCMRRSTV10,HaxthausenPK11,leuschelFFP11,FFGM13,ferrari2013metro,MNRST13,Gha14,JMNRST14,BosschaartQJG15,NGBPVMM16,BDG17,CLSQTL17,Gha17,LecomteDPM17,vu2017formal,BLW18,BBC18,BergerJLRS18,IliasovTLR18,MF18,Thai18,BBFL19,Abr20,AKJ20,CM20,DDPS20,HLKKNNSS20,MFTL20,TFLM20,BBL20,BBF0GMMPT20}. Notable cases are the usage of the B~method for developing the Line~14 of the Paris Métro and the driverless Paris–Roissy Airport shuttle~\cite{abrial2007formal}, the use of Simulink for formal model-based development of the metro control system of Rio de Janeiro~\cite{ferrari2013metro}, and the application of NuSMV to the ERTMS/ETCS European standard for railway control and management~\cite{ChiappiniCMRRSTV10}. 

Recently, a stream of literature also emerged on the evaluation and comparison of formal tools in railways. 
Mazzanti et al.~\cite{mazzanti2018towards} replicates the same railway design with different tools, informally describing their peculiarities. 
Haxthausen et al.~\cite{haxthausen16comparing} compares two methods for the verification of an interlocking system. Basile et al.~\cite{BBGFGS20} consider the differences between two formalisms and their associated simulation-based tools by applying them to an industrial railway project. Within this set of comparative  works, the H2020 European Project ASTRail (SAtellite-based Signalling and Automation SysTems on Railways along with Formal Method and Moving Block Validation)\,\footnote{\url{http://www.astrail.eu}}, foresees a structured analysis of the formal tools' landscape and the selection of the most appropriate ones for railways~\cite{FBMBFGPT19}. To address this goal, the project includes three main empirical activities: a systematic literature review, a systematic evaluation of tools, and an experimental assessment on a railway case study. 

As part of the ASTRail project, and building on a previous exploratory judgment study~\cite{FMBBF20}, this work reports on the \textit{systematic evaluation} of 13~formal tools for railway system design. We adapt the DESMET methodology for tool evaluation proposed by Kitchenham et al.~\cite{kitchenham1997desmet}. We first select a comprehensive set of evaluation features based on a brainstorming involving both formal methods and railway experts. Then, three assessors evaluate the tools and assign values to the features. To this end, they access the documentation, run the tools and perform trials to have an informed judgment. 
Multiple iterations and triangulation sessions are performed to ensure homogeneity in the assessment. 
As usability of the tools is considered as particularly relevant by railway practitioners according to recent surveys~\cite{BBFGMPTF18,BBFFGLM19}, a preliminary usability study is carried out involving subjects from a company. More specifically, the subjects assist to structured live demos of a subset of the tools, and fill in a usability test questionnaire. This evaluation setting  ensures a trade-off between the need to capture the viewpoint of railway experts, and the time that is inherently required to learn formal tools~\cite{steffen2017physics}. The data of our study---including a detailed description of evaluation features, evaluation sheets for the analysed tools, and data from the usability test---is shared to facilitate inspection, replication and extension~\cite{mazzanti2021data}.

Our work provides the following main contributions: i)~we establish a set of features to systematically evaluate formal tools for railways, which can be adapted to other domains; ii)~we assess the features on 13~different formal tools and share the tool evaluation sheets; iii)~we perform an initial comparative usability study of 7~of these tools involving railway practitioners; iv)~we debunk some common beliefs about formal tools, especially concerning usability and maturity. 


The rest of the paper is structured as follows. Sect.~\ref{sec:background} discusses background and related work. In Sect.~\ref{sec:design}, we illustrate the study design, while in Sect.~\ref{sec:results} we report the results. Sect.~\ref{sec:discussion} discusses implications. Sect.~\ref{sec:threats} reports threats to validity and Sect.~\ref{sec:conclusion} provides some final remarks.


\section{Background and Related Work}
\label{sec:background}


\subsection{The Railway Domain and the ASTRail Project}
\label{sec:railway}

The railway domain is characterized by a stringent safety culture, supported by rigorous development practices that are oriented to prevent catastrophic system failures and consequent accidents. The CENELEC EN~50128 standard for the development of software for railway control and protection systems~\cite{cenelec50128} considers formal methods as highly recommended practices when developing platforms of the highest Safety Integrity Levels (SIL-3/4). Given these premises, the railway domain has been a traditional playground for practical experimentation with formal methods and tools~\cite{fantechi2013twenty,woodcock2009formal,BGK18,GBP20}. 

The European Shift2Rail initiative\,\footnote{\url{http://www.shift2rail.org}} stimulates the development of safe and reliable technological advances to allow the completion of a single European railway area with the ambitious aim to \lq\lq double the capacity of the European rail system and increase its reliability and service quality by 50\%, all while halving life-cycle costs.\rq\rq\
Shift2Rail funds several projects, among which ASTRail (SAtellite-based Signalling and Automation SysTems on Railways along with Formal Method and Moving Block Validation). 
The project has the goal to introduce novel technologies within the railway industry, borrowing from the automotive and avionics domain. To this end, ASTRail studies the integration of satellite-based train positioning based on a Global Navigation Satellite System (GNSS), moving block distancing, and automatic train driving, with the support of formal methods. One of the objectives of the project is to identify the most suitable formal methods and tools to be used for railway system development. To this end, ASTRail considers three main steps: (1) a systematic literature review on formal methods and tools for railways, complemented with surveys with practitioners, and a review of research projects; (2) a systematic analysis of the most promising formal tools, based on the results of the previous activity; (3) a case study in which a selection of tools is used for the development of formal models of a railway signaling system that includes the aforementioned technologies. Results about the literature review and the surveys are reported in previous works~\cite{FBMBFGPT19,BBFGMPTF18,BBFFGLM19}. The current work presents the systematic analysis of a selection of tools. Before this analysis, we also  carried out a judgment study, involving \textit{formal methods experts}, in which 9 different tools were qualitatively analysed and specific strengths and weaknesses were identified~\cite{FMBBF20}.

\subsection{Formal Methods and Tools}

Formal methods are mathematics-based techniques useful for the specification, analysis and verification of  systems~\cite{wing1990specifier}. These methods are normally oriented to the development of mission-critical software and hardware, and are supported by software tools that can be used to model the system, perform automated verification, and, in some cases, produce source code and tests.  
The different formal techniques and tools that we consider can be partitioned into four broad families, namely \emph{theorem proving}~\cite{Duf91,RV01,bertot2013interactive}, 
\emph{model checking}~\cite{CGP99,BK08,CHVB18}, \emph{refinement checking}~\cite{gibson2014fdr3} and \emph{formal model-based development}\cite{FFGM13}\,\footnote{As we focus on system \textit{design}, we do not consider deductive verification or abstract interpretation, which are applied to source code.}. Here, we briefly mention some of the main characteristics of these different families and point to some representative tools.

\textit{Theorem proving} is an automated reasoning or deduction technique that aims 
to formally verify that some property holds in a system specification by means of semi-automated proofs that require interaction with the user. Roughly speaking, a theorem prover assists the user in demonstrating theorems over a system specification. The theorems can be related to some invariant that must hold in the specification or to the consistency of some specification refinement. Well-known theorem provers are Atelier~B~\cite{atelierb}, Coq~\cite{Ber08}, Isabelle~\cite{paulson1994isabelle} and PVS~\cite{OS08}.

\textit{Model checking} is a technique to verify that some desired property expressed with a declarative formalization, typically a temporal logic, is satisfied by a specification. It is common to distinguish between linear-time and branching-time model checking, depending on the type of temporal logic supported by the verification engine, and thus the way in which the desired properties can be expressed. 
The popularity of model checking is mainly due to its full automation. 
Traditionally, the model-checking problem is solved by generating and traversing the entire state space of the specification, composed with the property to be analyzed. This is the approach of tools such as ProB~\cite{leuschelFFP11}, SPIN~\cite{holzmann2004spin} and UMC~\cite{TERBEEK2011119}.

The full state-space exploration leads to the well-known state-space explosion problem, and several techniques have been developed to overcome this scalability issue, such as for example symbolic model checking~\cite{McM93}, supported by tools like NuSMV/nuXmv~\cite{CCCGR00,CCDGMMMRT14} and PRISM~\cite{KNP11}, which  applies also probabilistic model checking; statistical model checking~\cite{AP18,LLTYSG19}, applied by tools like UPPAAL~\cite{David2015}; and automata minimization techniques, which are supported by tools like CADP~\cite{CADP} and mCRL2~\cite{mCRL2}, both based on the formalism of process algebras.  

Other tools, such as TLA+~\cite{10.5555/579617} and SAL~\cite{10.1007/978-3-540-27813-9_45}, take a different approach and combine model checking and theorem proving. Theorem proving does not depend on the problem space, and this avoids the state-space explosion problem, although at the cost of reduced automation.

\textit{Refinement checking}~\cite{gibson2014fdr3} is a verification technique that consists in automatically assessing that a certain specification is a correct refinement of a higher-level specification. This is the main technique used by FDR4~\cite{gibson2014fdr3}, but it is supported also by other tools, such as ProB~\cite{leuschelFFP11}.  


\textit{Formal model-based development} tools, finally, focus on specification through graphical models, which can then be simulated, analyzed and verified by means of various techniques. Although these tools often embed model-checking capabilities to support verification, their main emphasis is on the modeling phase, and for this reason we classify them in a different family. Examples of such tools are Colored Petri Net (CPN) Tools~\cite{JKW07}, Modelica~\cite{modelica}, SCADE~\cite{10.1007/978-1-4020-6254-4_2} and Simulink~\cite{dabney2004mastering}.

\subsection{Related Work}

The different families of formal tools briefly listed above give an intuition of the wide landscape of choice available. 

On the other hand, empirical comparisons between tools, both for railways and other fields, are limited~\cite{FMBBF20,schlick2018proposal,steffen2017physics}. This, together with other issues largely discussed in the literature, such as the prototypical level of the tools, the skills required, the psychological barriers, tool deficiences, etc.~\cite{woodcock2009formal,davis2013study,GBP20,GM20} has seriously hampered the adoption of formal tools by industrial practitioners. 

 To address this issue, several efforts have been performed in the community of formal methods for system and software engineering. Early comparative works are those by Gerhart et al.~\cite{GCR94} and Ardis et al.~\cite{ACJMPSO96}. The former analyses the results of 12~case studies, and reports a lack of adequate tool support at the time of writing. The latter, instead, performs a critical evaluation of 7~formal specification languages, highlighting that maintainability of the models, especially in terms of modularity, is one of the most common issues to address. More recently, Abbassi et al.~\cite{ABDS18} qualitatively compares three specification languages, namely B, TLA+ and Dash---an extension of the Alloy language, and show specific language features that characterize each language. Overall, these works mainly focus on methods and specification languages, rather than tools. 

Well-known initiatives targeting a  comprehensive and comparable view of formal tools are: tool competitions (cf., e.g.,~\cite{bartocci2019toolympics}), mostly oriented to assess performance over shared benchmarks~\cite{garavel2019reflections}; the ETI initiative~\cite{steffen1997electronic}, aimed at creating an online service to experiment with different formal tools; the comprehensive study by Garavel and Graf~\cite{garavel2013}, a report of over 300~pages characterizing the landscape of available methods and associated tools; the recent expert survey on formal methods by Garavel et al.~\cite{GBP20}, a report of 67 pages on the past,
present, and future of formal methods and tools in research, industry, and education based on a survey among 130 high-profile experts in formal methods. 

In the railway domain, different surveys have been performed about formal methods and tools (cf., e.g.,~\cite{bjorner2003new,abrial2007formal,fantechi2013twenty,FWM13,flammini2012railway,boulanger2014formal}). 
 Bj{\o}rner~\cite{bjorner2003new} presents a first, non-systematic survey of formal methods applied to railway software. 
With a focus on the B~method, the book edited by Boulanger and Abrial~\cite{boulanger2014formal} discusses successful industrial usages, including railway experiences at Siemens and other companies.
Aspects of railway system development are covered by the book edited by Flammini~\cite{flammini2012railway}, where two entire chapters are dedicated to formal methods applications. New applications and future challenges related to the increasing complexity of railway systems are indicated in the reviews of Fantechi et al.~\cite{fantechi2013twenty,FWM13}. 
A special issue dedicated to formal methods for transport systems contains two experiences in the railway domain~\cite{BGK18}. 
A systematic literature review~\cite{FBMBFGPT19}, including 114~research papers on formal methods and railways, and two surveys with practitioners~\cite{BBFFGLM19,BBFGMPTF18}, were also recently published, highlighting the importance given by the railway industry to \textit{maturity} and \textit{usability} of formal tools. 

 Efforts in comparing formal tools have been performed also in railways. Specifically, Haxthausen et al.~\cite{haxthausen16comparing} compares two formal methodologies for interlocking system development. Mazzanti et al.~\cite{mazzanti2018towards} replicates the same railway design with multiple formal methods, and qualitatively discusses the peculiarities of each tool, namely CADP, FDR4, NuSMV, SPIN, UMC, mCLR2 and CPN Tools. Basile et al.~\cite{BBGFGS20} compares two formalisms and their associated simulation-based tools, among which UPPAAL, by applying them to a case study from an industrial railway project. During the ABZ 2018 conference, a case study track was specifically dedicated to a railway problem~\cite{hoang2018hybrid}. The specification provided by the organizers was modeled by different authors through different tools, including SPIN, Atelier~B and ProB, leading to independent publications (e.g.,~\cite{AKJ20,DDPS20,HLKKNNSS20,MFTL20}) in a dedicated special issue~\cite{BHRR20}. 
Finally, Ferrari et al.~\cite{FMBBF20} performs a judgment study 
considering 9~tools (all those used by Mazzanti et al.~\cite{mazzanti2018towards}, except for mCRL2 and CPN Tools), and produces a table with strengths and weaknesses, to guide the adoption of formal methods in railways. The study does not consider specific evaluation features, and takes more of a bird-eye  view on the tools.

\textit{Contribution.} Overall, none of the existing studies and initiatives, both in railways and in the other domains, 
performs a \textit{systematic} evaluation of formal tools.  The current work addresses this gap. Compared to previous contributions in railways~\cite{haxthausen16comparing,mazzanti2018towards,FMBBF20,hoang2018hybrid}, our work: 1)~is systematic and considers a comprehensive set of evaluation features; 2)~takes into account a larger set of tools; and 3)~is the first that presents a usability study with railway practitioners.  





\section{Research Design}
\label{sec:design}
Our study is oriented to provide a catalogue of the characteristics of available formal tools for railway system design, and achieve a more structured understanding of the field. 
To this end, we perform a systematic tool evaluation adapting the guidelines of the DESMET methodology by Kitchenham~\cite{kitchenham1997desmet}, according to the qualitative feature analysis paradigm. Specifically, we first select a set of relevant features, and then we consider existing documentation and perform tool trials to qualitatively evaluate the features with a sufficient degree of objectivity. Usability in terms of ease-of-use is evaluated through a usability study. The produced comparison table and the results of the usability study are used as a basis to compare the different tools and discuss research gaps as well as mismatches between railway designers' expectations as identified by previous work~\cite{BBFGMPTF18,BBFFGLM19}, and tools' functionalities and qualities. 
Although the DESMET methodology normally suggests a numerical scoring scheme for the evaluation features (i.e., qualitative values are mapped to numbers), we did not adopt this approach, as our goal is not to rank the tools, but to provide a comprehensive comparison without the need to establish winners and losers.

Our research focuses on the railway domain, given the long history of applications of formal methods in this field, and the demand for a more widespread knowledge dissemination towards practitioners and system designers~\cite{FMBBF20,schlick2018proposal}. 

\subsection{Research Questions}
Our overall research goal is to \textit{systematically evaluate formal tools for system design in the railway domain}. The goal is decomposed into the following research questions (RQs): 
\begin{itemize}
\item 
RQ1: \textit{Which are the features that should be considered to evaluate a formal tool?} This question aims at creating a taxonomy of features to evaluate the tools. To address this goal, we perform a brainstorming session of feature elicitation, followed by consolidation activities to come to a well-defined set of features and associated  values. 

\item 
RQ2: \textit{How do different tools compare with respect to the different features?} With this question, we want to highlight strengths and weaknesses of the tools according to the selected features. The 13~tools are individually evaluated by three assessors with the support of the tools' documentation, research papers, and through brief trials with simple draft models defined by the assessors themselves. The evaluation enables the identification of individual as well as collective weaknesses and strengths of the tools. For each tool, an evaluation sheet is produced.

\item 
RQ3: \textit{How do different tools compare with respect to usability?}
Usability is a particularly relevant macro-feature that needs to be assessed by different subjects to be properly evaluated. 
To compare the ease-of-use of the tools we perform a preliminary usability assessment with railway practitioners. The practitioners did not directly use the tools, but assisted to structured tools' live demos, and used the System Usability Score (SUS) test to evaluate them. This unorthodox usability assessment compensates for the inherent complexity of learning formal methods, and give a first intuition of the potential \textit{ease-of-use} the tools from the viewpoint of practitioners.
\end{itemize}

\subsection{Tools}
The 13~selected tools for the evaluation are 
CADP~(2020-g), 
FDR4~(4.2.7), 
NuSMV~(1.1.1), 
ProB~(1.9.3), 
Atelier~B~(4.5.1), 
Simulink~(R2020a), 
SPIN~(6.4.9), 
UMC~(4.8), 
UPPAAL~(4.1.4), 
mCLR2~(202006.0), 
SAL~(3.3), 
TLA+~(2)
and CPN Tools~(4.0).
The first ten were included as they are evaluated among the most mature formal tools for railways, according to Ferrari et al.~\cite{FMBBF20}. 
The last four were added to consider a more representative set of the different flavors of formal system modeling and verification available. Specifically, CPN Tools was included to consider Petri Nets, a widely used graphical formalism also for railway modeling~\cite{Thai18,BBFFGLM19}. mCLR2 is based on the algebra of communicating processes and supports minimization techniques, as does CADP, but mCRL2 is open source~\cite{mCRL2}. Finally, SAL and TLA+ are particular tools that integrate both theorem proving and model checking, and TLA+ is also used at Amazon~\cite{newcombe2014amazon}.
More information about the peculiarities of each tool is reported in the evaluation sheets that were produced for each tool~\cite{mazzanti2021data}. 
Further work can update the list of tools, using the features proposed in this paper as reference for assessment. 
The tools were tried in their free or academic license, depending on the available options. The commercial license was not purchased for any of the tools. 

\begin{table}[h]
\centering
\caption{\label{tab:participants}Characteristics and expertise of the study participants}
\setlength{\tabcolsep}{2pt}
\renewcommand{\arraystretch}{1.2}
\small{
\resizebox{\columnwidth}{!}{%
\begin{tabular}{|r|l|l|l|c|c|r|r|r|}
\hline
\multirow{3}{*}{\textbf{ID}} & 
\multicolumn{1}{c|}{\multirow{2}{*}{\textbf{Role in}}} &
\multicolumn{1}{c|}{\multirow{3}{*}{\textbf{Milieu}}} &
\multicolumn{1}{c|}{\multirow{3}{*}{\textbf{Main Function}}} &
\multirow{3}{*}{\textbf{Age}} &
\multirow{3}{*}{\textbf{Sex}} &
\multicolumn{3}{c|}{\textbf{Years of Experience in}} \\ \cline{7-9}
 & \multicolumn{1}{c|}{\multirow{2}{*}{\textbf{Study}}} & & & & & \multicolumn{1}{c|}{\textbf{Formal}} & \multicolumn{1}{c|}{\textbf{Railway}} & \multicolumn{1}{c|}{\textbf{FM in}} \\
 & & & & & & \multicolumn{1}{c|}{\textbf{Methods\,(FM)}} & \multicolumn{1}{c|}{\textbf{Industry}} & \multicolumn{1}{c|}{\textbf{Railways}} \\ \hline
1 & assessor & academic & workpackage leader & 39 & M & $>13$ & $3$ & $13$ \\ \hline
2 & assessor & academic & tool developer & 62 & M & $>20$ & $0$ & $9$ \\ \hline
3 & assessor & academic & researcher & 36 & M & $>\,~6$ & $0$ & $4$ \\ \hline
4 & expert & academic & group leader & 48 & M & $>15$ & $0$ & $9$ \\ \hline
5 & expert & academic & project leader & 66 & F & $>30$ & $0$ & $>25$ \\ \hline
6 & expert & academic & professor & 65 & M & $>30$ & $0$ & $>25$ \\ \hline
7 & expert & industry & system engineer & NA & M & $0$ & $>10$ & $0$ \\ \hline
8 & expert & industry & system engineer & 52 & M & $0$ & $>10$ & $0$ \\ \hline
9 & expert & industry & system engineer & 48 & M & $0$ & $>10$ & $0$ \\ \hline
10 & expert & industry & software developer & 43 & M & $0$ & $>10$ & $0$ \\ \hline
11 & expert & industry & product manager & NA & M & $0$ & $>10$ & $0$ \\ \hline
12 & expert & industry & system engineer & 48 & M & $0$ & $>10$ & $0$ \\ \hline
13 & expert & industry & innovation engineer & NA & M & $0$ & $>10$ & $0$ \\ \hline
14 & expert & industry & software developer & 45 & M & $0$ & $>10$ & $0$ \\ \hline
15 & expert & industry & innovation engineer & NA & F & $0$ & $3$ to $10$ & $0$ \\ \hline
\end{tabular}
}}
\end{table}

\subsection{Study Participants}
Table~\ref{tab:participants} summarizes the study participants' characteristics and expertise.

The main participants are the first three authors of the paper, who were involved in all the phases of the study. They are referred to as \textit{assessors}. The assessors 
are male academics, with 13~years (1st author), 9~years (2nd author) and 4~years (3rd author) of experience in applications of formal methods to the railway industry, but with complementary expertise: semi-formal methods, classical model checking, and probabilistic and statistical methods and tools, respectively. 

The association between tools and assessors was based on their expertise, to have an authoritative analysis. It is worth noting that the 2nd author has over 25~years of experience in formal methods, as well as hands-on experience on formal tool development, while the 1st author has 3~years of experience with applying semi-formal methods in the railway industry. The adopted DESMET methodology is inspired by systematic literature reviews~\cite{K04}, in which the authors are normally involved in the evaluation, and, as in our case, aims to reach objectivity through the selection of appropriate evaluation features, triangulation, and cross-checking between the authors. 

The additional participants included three researchers (referred to as \textit{academic experts}) and 9~railway practitioners (\textit{industry experts}). One of the researchers is the last author, a male academic that has more than 15~years of experience with multiple formal methods, with a specific focus on the application of model-checking tools, among which probabilistic and statistical variants, recently also to railway problems. The other two academics (one male and one female) each have more than 25~years of experience in the application of formal methods to railways. Finally, the 9~practitioners, of which one female, have in general more than 10~years of experience in railways, but no prior experience in the application of formal methods. 

\subsection{RQ1: Feature Selection}
The features were elicited with a collaborative approach inspired by the KJ method~\cite{scupin1997kj}. A 3~hour workshop was organized involving 8 participants: the assessors, the academic experts, and two of the industry experts (ID 8 and 9 in Table~\ref{tab:participants}). Participants were given 5~minutes to think about relevant features that should be considered when evaluating a formal tool for railway system design. Each participant wrote the features that they considered relevant in a sheet of paper that was not made visible to the others. Then, the moderator asked the participants to list their features and briefly discuss them. When the explanation given by the feature proponent was not clear, the others could ask additional questions to clarify the meaning of the feature. The moderator reported each feature in a list that was made visible to all participants through a projector. When a feature was already mentioned, it was not added to the list. At the end of the meeting, the assessors homogenized the feature names. For each feature, the assessors jointly defined the possible values, as well as a systematic way to assess it. Features that could not be evaluated with a sufficient degree of objectivity, and that required specific experimental evaluations (i.e., notation readability, resource consumption, learning curve, and \textit{scalability}~\cite{steffen2017physics}) were excluded from the analysis. The final list of features and possible values was later cross-checked by the other participants. A reference evaluation document was redacted with all the information to be included for each tool. 

\subsection{RQ2: Feature Evaluation}
The three assessors performed the systematic feature evaluation.  
The assessors worked independently on a subset of the considered tools, and they produced an evaluation sheet for each tool, based on the reference document. The assessors worked on the tools that they were more familiar with, according to their previous experiences, so to give a more informed judgment. Specifically, one assessor (3rd author) used UPPAAL and Atelier B; one used Simulink 
(1st author); the assessor with more experience in formal methods (2nd author) used the remaining tools. 
To perform the evaluation, the assessors followed a structured procedure: 1)~install and run the tool; 2)~consult the website of the tool, to check the official documentation; 3)~opportunistically search for additional documentation to identify useful information to fill the evaluation sheet; 4)~refer to the structured list of papers on formal methods and railways published in the previously referred literature review~\cite{FBMBFGPT19}\,\footnote{The list of 114~papers, and associated categories, can be accessed at https://goo.gl/TqGQx5.} (cf.\ Sect.~\ref{sec:railway}), to check for tools' applications in railways; 5)~perform some trials with the tools to confirm claims reported in the documentation, and assign the value to those features that required some hands-on activity to be evaluated; 6)~report the evaluation on the sheet, together with the links to the consulted documents and papers, and appropriate notes when the motivation of some assignment needed clarification. 

In subsequent face-to-face meetings, the assessors challenged each others' choices, and asked to provide motivations and evidence for the assigned values. The evaluation sheets were further revised by all three assessors to align visions, and balance judgments. The process was incremental, and carried out across several months. In total, six meetings of~1 to~2 hours were carried out, and the evaluation sheets were used as living documents during these meetings. Finally, a summarizing table was produced to systematically compare the tools. The final evaluation sheets were part of a deliverable of the ASTRail project, and were thus subject to external review. Furthermore, the last author of this paper  cross-checked the table with the evaluation sheets. The sheets, together with a detailed description of features and assessment criteria, are shared in our repository~\cite{mazzanti2018towards}. 

\subsection{RQ3: Usability Evaluation}
This section outlines the methodology adopted for the usability evaluation of the tools performed with railway experts. First, a set of models of the same sample system was developed by the assessors using 7~different tools, namely Atelier~B, NuSMV, ProB, Simulink, SPIN, UMC, and UPPAAL. 
The other tools were not included as, from the feature evaluation, they were considered to require \textit{advanced} mathematical background to be understood (i.e., CADP, FDR4, mCLR2, SAL, or TLA+, cf.\ Fig.~\ref{fig:table}) or they were known from the literature to be inadequate to handle industry-size problems (this is the case of CPN Tools~\cite{mazzanti2018towards}). One exception is Atelier~B, which, although requiring advanced background, is one of the few tools already used in the development of real-world railway products (cf.\ ``Integration in the CENELEC process'', Fig.~\ref{fig:table}). 

Each assessor worked only on a subset of the tools, depending on their skills, as in the feature evaluation phase. The models were used to showcase the different tools, and evaluate their usability from the point of view of the 9 industry experts. A usability assessment in which the experts would directly interact with the tools was not considered reasonable, due to the skills required to master the tools, and due to time constraints. Therefore, the three assessors showed the different characteristics of the tools in a 3~hour meeting with the experts, using the developed models as a reference. The experts were already familiar with the sample system considered, i.e., the moving block system, which was also used as reference by other works in the literature~\cite{hoang2018hybrid,BHRR20,FMBBF20}.  The assessors asked the experts to evaluate the usability of each tool based on their first impression. It should be noticed that, although unorthodox, this approach is meaningful. In practical cases, formal tool users need to be tool experts~\cite{mazzanti2018towards,steffen2017physics}, and railway engineers will have to interact with them. This requires the engineers to be able to understand the models and make sense of the results, but not to be proficient with the formal tools. 

The meeting was performed as follows:

\begin{enumerate}
\item \textbf{Introduction:} an introduction was given to recall the main principles of the considered system. This served to provide all participants a uniform perspective on the system that they were going to see modeled.
\item \textbf{Tool Showcase:} each tool was presented by an assessor in a 15 minutes demo, covering the following aspects: 
1. \textit{General structure of the tool:} the presenter opens the tool, and provides a description of the graphical user interface (if available);
2. \textit{Elements of the model:} the presenter opens the model, describes its architecture, and navigates it;
3. \textit{Elements of the language:} minimal description of the modeling language constructs, based on the model shown;
4. \textit{Simulation features:} a guided simulation is performed (if supported); 5. \textit{Verification features:} description of the language used for formal verification, and presentation of a formal verification session with counter-example (if supported).
\item \textbf{Usability Evaluation:} after the presentation of each tool, a usability questionnaire for the tool (see below) is filled by the experts. 
\end{enumerate}

To evaluate the usability of the tool, we use a widely adopted usability questionnaire, the System Usability Scale (SUS) developed by Brooke~\cite{brooke1996sus,Bro13}. We preferred SUS over other questionnaires like QSUC, QUIS, and USE (cf.~\cite{pourali2018empirical} for pointers), since the questions were considered appropriate for our context. Some questions from the original SUS had to be tailored to our evaluation. The final questionnaire was:

\begin{enumerate}
\item I think that I would like to use this tool frequently.
\item I found the tool unnecessarily complex.
\item I thought the tool was easy to use.
\item I think that I would need the support of a technical person to be able to use this tool.
\item I found the various functions in this tool were well integrated.
\item I thought there was too much inconsistency in this tool.
\item I would imagine that most people with industrial railway background would learn to use this tool very quickly.
\item I imagine that the tool would be very cumbersome to use.
\item I imagine that I would feel very confident using the tool.
\item I imagine I would need to learn a lot of things before I could get going with this tool.
\end{enumerate}

Answers are given in a 5-points Likert Scale, where 0 = Completely Disagree; 1 = Partially Disagree; 2 = Undecided; 3 = Partially Agree; 4 = Agree. 
To calculate the SUS score, we follow the guidelines of Brooke~\cite{brooke1996sus}. 
Overall, the SUS Score varies between 0 and 100, with the following interpretations for the scores, based on the work of Bangor et al.~\cite{bangor2008empirical}: 100~=~Best Imaginable; 85~=~Excellent; 73~=~Good; 52~=~OK; 39~=~Poor; 25~=~Worst Imaginable. 

\section{Threats to Validity}
\label{sec:threats}


\emph{Construct Validity}. The set of considered tools does not represent the complete universe of formal tools. However, the selection rationale was motivated by other works in railways~\cite{FMBBF20}, and by the need to have representatives of different families (cf.\ Sect.~\ref{sec:design}). Overall, there is a bias towards model-checking tools with respect to theorem proving ones, but this also occurs in the railway literature, where the preference for this technique is rather common~\cite{FBMBFGPT19}. It is worth noting that we did not purchase commercial licenses for those products offering them. This should not affect our evaluation, as academic licenses appear to support the same features. 

The different features, as well as the possible values, were defined by a limited group of persons. However, the group represents both academic and railway industry viewpoints, and several triangulation activities were carried out to ensure clarity, a sufficient degree of completeness and a uniform interpretation. 
We moreover made an effort to define sufficiently objective features. In fact, we argue that the partially subjective features are a minority (namely, easy to install, quality of documentation, and complexity of license management).

The means that was adopted to evaluate usability, i.e., the SUS questionnaire, is widely used and has been proven effective~\cite{bangor2008empirical}. On the other hand, the presented usability evaluation is based on a demo showcase of the tools and 
different results may be obtained with a direct evaluation by railway experts. Given the relatively high number of tools and the resources required for such a direct evaluation by railway experts, we opted for a pragmatic approach that provides an indication of the potential usability of the tools. 

\emph{Internal Validity}.
The evaluation of the tools may have suffered from subjectivity. To limit this problem, triangulation activities were performed to align the evaluation, and the assessors were required to report appropriate evidence in the evaluation sheet, when some judgment required explicit justification. For some features, the value also depends on the available evidence that comes from the literature and from the websites of the tool, and may not reflect reality. To mitigate these aspects, different sources of information, and practical tool trials have been considered. 

In the usability study, the researchers could have biased the audience towards a certain tool, based on their preferences. To mitigate this threat, the researchers rehearsed the tool showcase before the evaluation with the participation of one of the academic experts, and provided mutual recommendations. Therefore, we argue that the tools were presented in a quite uniform manner. 

Another issue is related to the number of subjects involved, i.e.~9, which may be regarded as a limited sample. However, for different methods it has been shown that a group of 10$\pm$2~evaluators can identify $80\%$ of the usability problems and that ``optimal  sample  sizes  of `10$\pm$2’ can  be applied  to a  general  or  basic  evaluation  situation,  for  example,  just  basic training  provided  to  evaluators  and  a limited  evaluation  time  allowed''~\cite{hwang2010number}. Therefore, we argue that, although limited, the sample size is in line with the samples typically used in usability testing.

\emph{External Validity}. 
The usability evaluation, as well as the preceding activities, involved railway experts with multiple roles, i.e., system engineers, developers, and managers, and with different degrees of experience in railways, although more than 10 years in the vast majority of the cases. This covers a large spectrum of perspectives. On the other hand, the participants were all from the same railway supply company, and this may limit the results. However, it is worth noticing that railway companies follow standardized processes and, in addition, railway systems are well-established and even standardized in some cases (e.g., ERTMS~\cite{FHAB17,hoang2018hybrid}). 
Moreover, the railway industry is an oligopoly and according to experts from the Shift2Rail project NEAR2050 (Future challenges of the railway sector)\footnote{\url{http://www.near2050.eu}} ``the rail supply industry will continue to be an oligopoly, but core rail system knowledge will have moved to IT departments or IT companies''~\cite{NEAR2050}.
Therefore, we argue that the perspectives of companies may be similar to each other and backgrounds and practices are comparable between companies, thus suggesting a sufficient degree of external validity of our results.

\section{Execution and Results}
\label{sec:results}

\subsection{RQ1: Feature Selection}

The feature selection process led to the identification of 33 features, which have been hierarchically grouped into \textit{functional}, \textit{language expressiveness}, and \textit{quality} features, with 8 categories in total. Below, we report categories, features and possible values with evaluation criteria. 



\paragraph{Functional Features} Functional aspects have been organized in two categories: development functionalities and verification functionalities.

\textbf{Development Functionalities.} 

\begin{itemize}
\item\textit{Specification, modeling:} specifies whether the model can be edited graphically (GRAPH), in some textual representation (TEXT) inside the tool or whether the model is just imported as a textual file (TEXTIM).
\item \textit{Code generation:} indicates if the tool supports automated code generation from specifications (YES, NO).
\item \textit{Documentation and report generation:} the tool supports automated generation of readable reports and documents (YES), allows the user to produce diagrams or partial reports that can be in principle included in official documentation (PARTIAL) or does not generate any usable documentation (NO).
\item \textit{Requirements traceability:} indicates if it is possible to trace requirements to the artifacts produced with the tool (YES, NO).
\item \textit{Project management:} specifies if the tool supports the management of a project, and the GUI-based navigation of its conceptual components (YES, NO). 
\end{itemize}

\textbf{Verification Functionalities.} 
\begin{itemize}
\item \textit{Simulation:} indicates whether a model is executable, so that simulation is possible. The simulation could be either graphical (GRAPH) or textual (TEXT), a mix of the two (MIX) or absent (NO).
\item \textit{Formal verification:} the type of formal verification supported can be linear-time model checking (MC-L), branching-time model checking (MC-B), observer-based model checking (MC-O), theorem proving (TP) or refinement checking (RF). 
\item \textit{Scalability approach:} indicates the type of approach adopted to verify large scale models. Can be on-the-fly model checking (FLY): the state is generated on demand; Partial order reduction (POR): exploitation of symmetries in the state space; Parallel computation (PAR): parallel computation distributed on more hosts; Bounded Model Checking (BMC): state-space exploration up to a certain depth; Symbolic Model Checking (SYM): compact state space representation; SAT/SMT constraint solving and theorem proving (SCT): avoid explicit reasoning on the state space; Statistical Model Checking (SMC): avoid full state-space generation using simulations and provide an approximate solution; Compositionality and minimization (COM): divide the problem into smaller subproblems; No technique (NO). 
\item \textit{Model-based testing:} support for automatically derived testing scenarios (YES, NO).
\end{itemize}

\paragraph{Language expressiveness} This group of features collects technical aspects related to the main modeling language made available by the tool.

\begin{itemize}
\item \textit{Non-determinism:} evaluates if non-determinism is expressible. In particular, whether the language allows internal non-deterministic system evolution (INT) or external choices associated to inputs or trigger-events allow the expression of non-determinism (EXT). 
\item \textit{Concurrency:} evaluates if and how concurrency aspects can be modeled. The model can be constituted by a set of asynchronously interacting elements (ASYNCH), synchronous elements (SYNCH), both ((A)SYNCH) or just one element (NO).
\item \textit{Temporal aspects:} considers if modeling language supports the notion of time (YES, NO).
\item \textit{Stochastic or probabilistic aspects:} evaluates if it is possible to model aspects related to randomness, such as for example stochastic delays (YES, NO).
\item \textit{Modularity of the language:} evaluates how the architecture of the model can be structured in the form of different hierarchically linked modules. The cases are: the tool allows the user to model in a hierarchical way, and the partitioning of the model into modules (HIGH); the tool allows the partitioning of the model into modules, but without a notion of hierarchy (MEDIUM); modules are supported, but the they have no way to interact, neither by messages nor by shared memory (LOW); modules not supported (NO).
\item \textit{Supported data structures:} the language  supports numeric types, but no composite expressions (BASIC) or it has complex expressions like sequences, sets, and array values (COMPLEX).
\item \textit{Float support:} indicates if floating-point numbers are basic data types (YES, NO).
\end{itemize}

\paragraph{Quality Features}
Quality aspects are organized into six categories: tool flexibility, maturity, usability, company constraints, and domain-specific criteria.

\textbf{Tool Flexibility.} 
\begin{itemize}
\item \textit{Backward compatibility:} indicates to which extent models developed with previous versions of the tool can be used in the current version. Cases are: the vendor guarantees that legacy versions of the models can be used in the current version of the tool or the future availability of legacy versions of the tool (YES); the tool is open source, the input language is stable and \textit{de facto} standard or there is evidence of interest in preserving backward compatibility (LIKELY); the tool is not open source and the provider does not show evidence regarding backward compatibility, even if the language is rather stable and a \textit{de facto} standard (MODERATE); source code is not available, input format is not stable, and no information is available from the vendor (UNCERTAIN).
\item \textit{Standard input format:} evaluates if the 
the input language is based on a language standardized by an international organization (STANDARD); the input language is open, public, and documented (OPEN); the structure of the model specifications is easily accessible, but not publicly documented (PARTIAL); the internal structure of the model specification is hidden (NO).
\item \textit{Import from or export to other tools:} evaluates whether the tool provides several import/export functionalities (HIGH); the tool has a standard format used by other tools or exports to some other formats (MEDIUM); the tool does not have import/export functionalities (LOW).
\item \textit{Modularity of the tool:} evaluates if the tool includes different modules and packages. Values are: the tool is composed of many modules that can be loaded to address different phases of the development process (HIGH); the tool offers multiple functionalities, but not in the form of loadable modules (MEDIUM); the tool offers a limited number of functionalities in a monolithic environment (LOW).
\item \textit{Team support:} specifies support for  collaborative model development (YES, NO).
\end{itemize}

\textbf{Maturity.} 
\begin{itemize}
\item \textit{Industrial diffusion:} the website of the tool reports multiple (HIGH), a few (MEDIUM), or no (LOW) cases of industrial usage; 
\item \textit{Stage of development:} the tool is a stable product with a long history of versions (MATURE); the tool is recent but with a solid infrastructure (PARTIAL); the tool is a prototype (PROTOTYPE).
\end{itemize}

\textbf{Usability}---in the broad sense of ISO 9241-11:2018~\cite{standard2019ergonomics}.
\begin{itemize}
\item \textit{Availability of customer support:} considers if reliable customer support can be purchased for maintenance and training (YES); free support is available in the form of bug reports and forums (PARTIAL); or communications channels need to be established between producers and users to have support (NO).
\item \textit{Graphical user interface:} the tool has a well-designed and powerful GUI (YES); a user-friendly GUI exists, but it does not cover all the tool functionalities in a graphical form (PARTIAL); a GUI exists, but not particularly powerful (LIMITED); the tool is command line (NO).
\item \textit{Mathematical background:} this feature aims at giving an idea of how easy it is to learn the tool for an electronic or computer engineer (i.e., the typical railway practitioner). Cases are: the tool does not require particular logical/mathematical skills  (BASIC); the tool requires the knowledge of temporal logic (MEDIUM); the tool requires the knowledge of theorem proving or process algebras (ADVANCED).

\item \textit{Quality of documentation:} the documentation is extensive, updated and clear, includes examples that can be used by domain experts, it is accessible and navigable in an easy way (EXCELLENT); the documentation is complete but offline, and requires some effort to be navigated (GOOD); the documentation is not sufficient or not easily accessible (LIMITED).
\end{itemize}

\textbf{Company Constraints.} 
\begin{itemize}
\item \textit{Cost:} the tool is available under payment only  (PAY); free under certain conditions (e.g., academic) and moderate cost for industrial use (MIX); the tool is free (FREE).
\item \textit{Supported platforms:} indicates possible platforms supported by the tool  (e.g., Windows, MacOS, Linux, or ALL three).
\item \textit{Complexity of license management:} the tool is free for commercial use, and no license management system is required (EASY); the tool offers academic and commercial licenses, both upon payment, limited effort was required to handle the academic license, and adequate information is provided for the commercial one (ADEQUATE); the tool has a free and a purchasable version, limited problems were encountered when trying the free version, but limited information is provided in the website about the licensing system of the other one (MODERATE); several problems were encountered with the license management system (COMPLEX).   
\item \textit{Easy to install:} the tool requires little or no external components (YES); the tool installation depends on external components or the installation process is not smooth (PARTIAL); the installation can interfere with the customer development environment (NO).
\end{itemize}

\pagebreak\textbf{Domain-specific Criteria.} 
\begin{itemize}
\item \textit{CENELEC Certification:} the tool is certified according to the CENELEC norm (YES); the tool includes a CENELEC certification kit or it is certified according to other safety-related norms like DO178C~\cite{rierson2017developing} (PARTIAL); none of the above (NO).
\item \textit{Integration in the CENELEC process:} estimates how easy it is to integrate the tool in the existing railway life-cycle as described by the CENELEC norms. Cases are: in the literature review from~\cite{FBMBFGPT19} or in the tool documentation evidence was found of tool usage for the development of railway products according to the CENELEC norms (YES); evidence was found of the usage of the tool in railways, but no CENELEC products developed with the tool (MEDIUM); no usage in railways was found (LOW).
\end{itemize}

\begin{figure*}
\centering
\includegraphics[width=\textwidth]{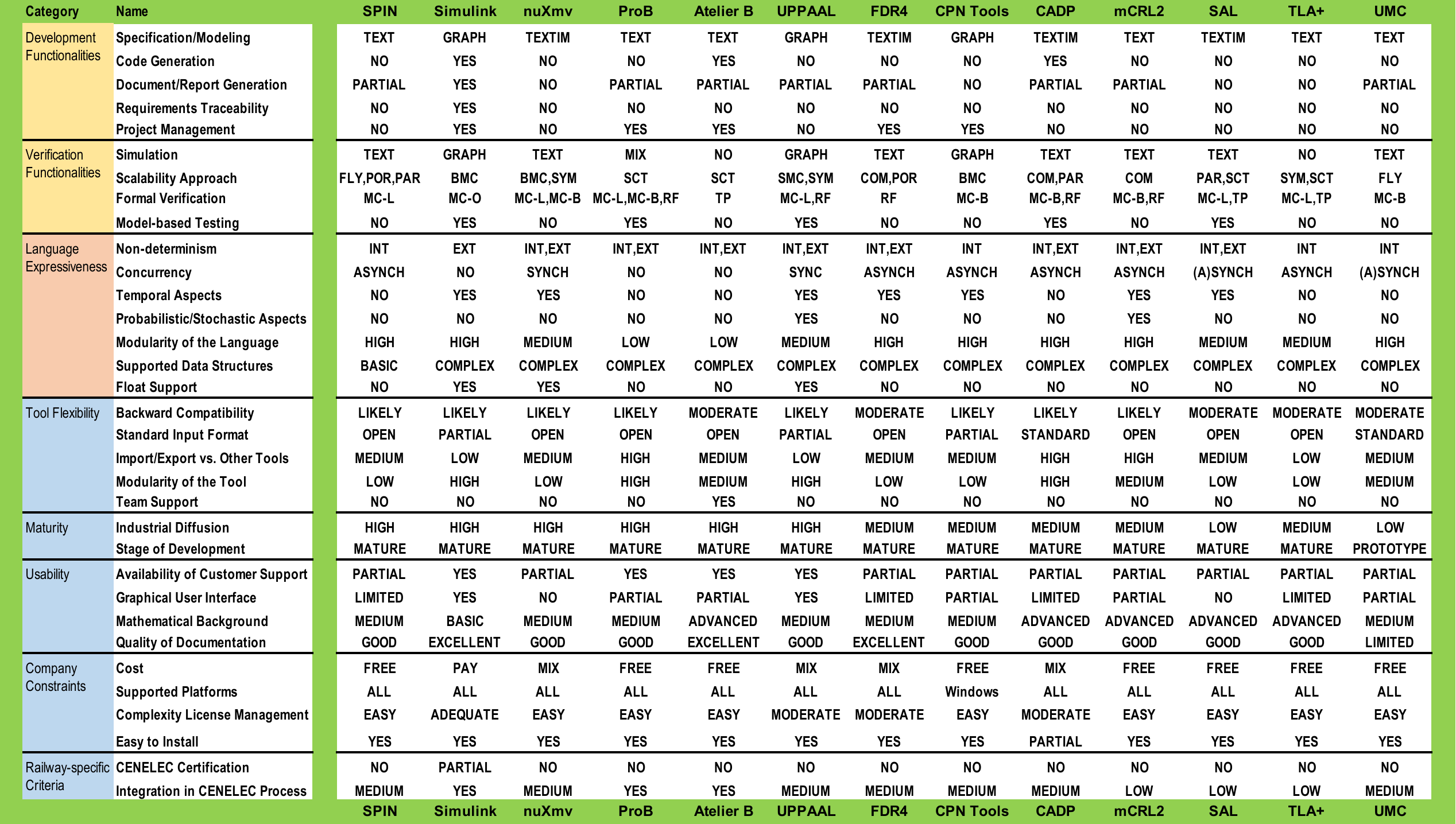}
\caption{Evaluation table}
\label{fig:table}
\end{figure*}

\subsection{RQ2: Feature Evaluation}
Figure~\ref{fig:table} reports the table resulting from the feature evaluation activity.
The reader is invited to consult the sheet of each tool to have  clarifications on the judgments provided. Here we summarise the most evident trends, and contrast them with existing literature on desired features of formal tools in railway as indicated by recent surveys~\cite{BBFGMPTF18,BBFFGLM19,FBMBFGPT19}, and limitations and barriers for formal methods adoption (cf.~\cite{ameur2010toward,davis2013study,garavel2019reflections,steffen2017physics,nyberg2018formal,atlee2013recommendations,schlick2018proposal,lamsweerde2000formal}).  

\textbf{Development Functionalities.} Concerning development functionalities, we observe that the majority of the tools are based on textual specifications of the models (TEXT and TEXTIM) and, with the exception of Simulink, most of them support only a limited subset of the other complementary functionalities, such as code generation and requirements traceability. Interestingly, the project management feature, common in any IDE for software development, is  only available in Simulink, ProB, and Atelier~B. 

\textit{Observations.} The tools appear to give limited  relevance to development functionalities. This, to our knowledge, was not observed by other authors discussing limitations of formal tools (cf.~\cite{davis2013study,garavel2019reflections,steffen2017physics,nyberg2018formal}). Traceability in particular is regarded as one of the most relevant functionalities by railway stakeholders~\cite{FBMBFGPT19}, who need to ensure that all the artifacts of the process are explicitly linked. The scarce support for traceability is therefore a relevant pain point. 
Concerning the type of specification language supported (graphical vs.\ textual), some railway stakeholders may in principle prefer graphical languages~\cite{FFGM13}, while few tools support them. However, Ottensooser et al.~\cite{ottensooser2012making} have empirically shown that a pictorial specification is not necessarily more understandable than a textual one, and code-like models may be easier to maintain. Therefore, the need for graphical languages---and the understanding of specifications by all stakeholders---could be conflicting with the need for flexibility. Regardless of the language format, a project management feature is needed to manage complex industrial models. 

\textbf{Verification Functionalities.} Verification functionalities, including model-based testing and simulation, are supported by a larger number of tools. However, specific strategies in terms of formal verification and scalability approaches are adopted by each platform, and this suggests a rather wide difference in terms of types of properties that can be verified on the models by each tool. 

\textit{Observations.} The difference between approaches is justified by the academic origin of most of the tools, which were primarily used to implement novel verification techniques, as observed by Garavel and Mateescu~\cite{garavel2019reflections}. This has an impact on the degree of experience required to master a tool. Several authors (cf.~\cite{mazzanti2018towards,steffen2017physics,garavel2019reflections}) highlighted that the differences in terms of verification strategies, together with the wide range of optimization options made available to address scalability issues, require users to be experts in the tool to successfully verify large designs. The presence of several tools supporting model-based testing suggest that tool developers appear to be aware of the need for  \textit{complementarity} between testing and formal verification, which are not interchangeable activities. 



\textbf{Language Expressiveness.} In terms of language expressiveness, the variety of feature value combinations, and therefore the specificity of each modeling language, is also quite evident, as for verification functionalities. In other terms, each tool is somewhat \textit{unique}, both in terms of specification and in terms of verification strategies. 
Probabilistic aspects and floats have in general limited support. Only UPPAAL and mCLR2 allow the expression of probability, and floats are native types only for Simulink, nuXmv, and UPPAAL. 

\textit{Observations.} The uniqueness of each language shows that after twenty years, the issues of \textit{isolation} of formal specification languages pointed out by Van~Lamsweerde~\cite{lamsweerde2000formal} have yet to be solved. 
The limited support for floats suggests that, in most of the cases, tools are oriented to designing models that do not go down to the expressiveness of source code in terms of numerical data representation. Abstracting from details is nevertheless one of the principles of systems modeling, and other tools oriented to static analysis should be used to deal with errors arising from floating point numerical types~\cite{moscato2017automatic}. 

We note that probabilistic features are traditionally associated to specialized tools for the analysis of performances and other dependability aspects~\cite{hartmanns2015quantitative,katoen2016probabilistic}. In this paper, we consider tools oriented towards system design, and this is the reason why this feature has limited support.

\textbf{Tool Flexibility.} Tool flexibility sees a major weakness in the team support feature, with Atelier B as the only tool including it. Only Simulink, ProB, UPPAAL, and CADP can be regarded as flexible toolboxes (cf.\ ``Modularity of the Tool'' = HIGH), and only ProB, CADP, and mCLR2 are open to different formats with several import/export functionalities. This suggests that in many cases the tools are independent ecosystems, and their integrated usage may be complicated.

\textit{Observations.} This issue also observed by other authors~\cite{garavel2019reflections,steffen2017physics}, who pointed out the high degree of specialization of languages and tools. Combining at least a subset of the tools and integrating them in a coherent process is extremely relevant. Indeed, as observed by previous work~\cite{FMBBF20,gleirscher2019new}, and given the ``uniqueness'' of each tool, the needs of an industrial process cannot be fulfilled by one platform only. 

\textbf{Maturity.} The majority of the tools has a mature stage of development, and diffusion appears to be medium-high. 

\textit{Observations.} The high degree of industry-readiness observed is in contrast with the observation of previous work (cf., e.g., ~\cite{atlee2013recommendations,woodcock2009formal}) reporting  that the low maturity and prototypical level of many tools are among the main obstacles for formal methods adoption. As maturity is, by far, recognized as the most relevant quality attribute that a formal tool should have to be applied in the conservative world of railways~\cite{BBFGMPTF18,BBFFGLM19}, we argue that low maturity can be considered as a false barrier for formal methods adoption.  

\textbf{Usability.} Usability aspects appear, at first glance, as major pain points, with limited customer support, limited GUI and the need for a medium to advanced mathematical background. 

\textit{Observations.} Ease of learning, which is the second-most desired feature~\cite{BBFGMPTF18,BBFFGLM19}, is notoriously a problem for formal tools~\cite{davis2013study,bjorner201440}, and it is complicated by the decreasing mathematical competence of engineers over the years~\cite{bjorner201440}. In this sense our work confirms the literature. On the other hand, it also highlights that, to address this problem, not only engineering curricula should be enriched, and better GUIs should be developed, but also more effort should be dedicated to customer support, which is acceptable only for Simulink, ProB, Atelier~B, and UPPAAL.  

\textbf{Company Constraints.} 
Company constraints are in general fulfilled, with several platforms supported and ease of install and license management. This confirms the maturity and industry-readiness of the majority of the tools---cf.\ ``Maturity'' for related observations. 

\textbf{Railway-specific Criteria.} 
Railway-specific criteria are not fulfilled, with MEDIUM/LOW easiness in integrating a tool in the CENELEC process, and no certified tool. However, three tools, namely Atelier~B, ProB, and Simulink have been used to develop railway products, indicating that their integration in the CENELEC process is at least possible.

\textit{Observations.} The issue of CENELEC integration is regarded as the third most desired quality feature by railway practitioners~\cite{BBFGMPTF18,BBFFGLM19}. It appears that a major pain point that industry may face is that practitioners do not have guidance to accommodate these tools within their industrial processes, and how to combine the tools together in a flexible way. This was also observed in other domains, such as  aerospace~\cite{davis2013study}. 

\subsection{RQ3: Usability Evaluation}
 
\begin{figure}
\centering
\includegraphics[width=\columnwidth]{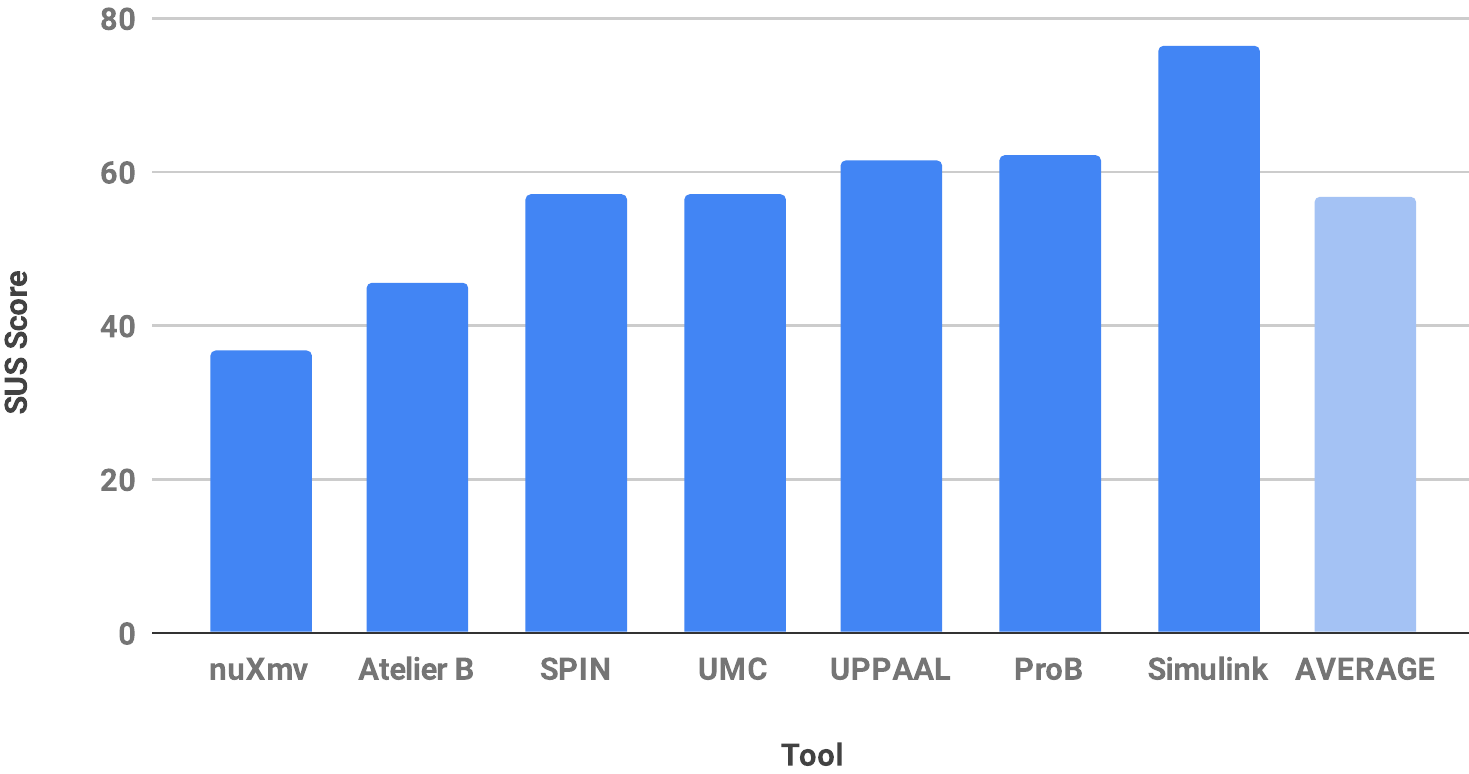}
\caption{SUS scores for the different tools}
\label{fig:sus}
\end{figure}

Figure~\ref{fig:sus} presents the results of the SUS questionnaire. The tool that clearly stands out as being considered the most usable is Simulink (SUS Score = $76.39$). This is a formal model-based development tool, with appealing, effective GUIs, powerful languages and simulation capabilities. 
Simulink is followed by three other tools with acceptable GUIs, but with widely different capabilities, namely ProB (SUS Score = $62.22$), UPPAAL ($61.67$), and UMC ($57.22$). ProB and UMC allow the user to model in textual form, but present the results of the simulation also in graphical form. Instead, the UPPAAL language is entirely graphical, and presents a graphical simulation style that recalls message sequence charts, which are well known by railway practitioners. 
Finally, SPIN (SUS Score = $56.94$), Atelier~B ($45.56$), and nuXmv ($36.67$), with some differences, are considered among the least usable tools. Although SPIN is a command-line tool, with only a limited GUI, its scores are higher than Atelier~B. This can be explained considering the following observations discussed with the participants: (a)~SPIN uses a modeling language that is very similar to the C language, and therefore was considered familiar by the participants, who, in turn, gave higher scores; (b)~Atelier~B uses a refinement-based theorem-proving approach, which requires advanced skills to be mastered. 

When computing the \textit{average} SUS Score, we obtain $56.67$, which is between OK and Good~\cite{bangor2008empirical}. Hence, the general usability of the tools can be considered acceptable. 


%

\section{Summary and Implications}
\label{sec:discussion}

Considering the inherent limitations of the DESMET methodology~\cite{kitchenham1997desmet}, our \textbf{take-away messages} are as follows:
\begin{framed}
\begin{enumerate}
    \item Many of the formal tools lack support for development features and process-integration aspects;
    \item Most of the formal tools are independent ecosystems, with unique, non-standard languages and specialized verification capabilities;
    \item Formal tools are mature, as highly desired by railway industry~\cite{BBFGMPTF18,BBFFGLM19};
    \item Most usability aspects appear to be low in principle, but, when the formal tools are assessed by practitioners, usability is considered acceptable. 
\end{enumerate}
\end{framed}
In the following, we discuss the main implications of our results. Our goal is to foster the debate about the adoption of formal tools in safety-critical software engineering, and therefore we also indulge in expressing personal opinions.  

\textit{Implications for Tool Developers.} The majority of the considered formal tools lack support for process-related aspects, including development functionalities such as traceability and document/report generation, and no evidence of their integration in the railway process. With some differences, exceptions are Simulink, Atelier~B, and ProB. Tool developers are encouraged to give more relevance to process-related aspects and enrich their tools with additional development functionalities. Furthermore, they are encouraged to invest in customer support and consultancy: companies need to be accompanied in the introduction of novel tools, and tailored recommendations need to be provided on how to best adapt their processes. Making their tool more interoperable/standardized is also suggested, as  importing from known formats could in principle  facilitate the transition of users to their tool.  

\textit{Implications for Researchers.} 
Tool certification, or qualification according to the norms, is regarded as a relevant problem by industrial practitioners, as also observed by Garavel~\cite{garavel2019reflections} and Mazzanti~\cite{mazzanti2018towards}. However, none of the tools considered here is CENELEC certified (the formal model-based development tool SCADE is). 
Nevertheless, if a formal tool does not go down to code generation, 
we argue that tool qualification is not a radical issue. Railway systems get certified also without using formal design tools, as these are highly recommended but not \textit{required}~\cite{nyberg2018formal}. We argue that evidence should be provided in terms of \textit{cost-benefit} analyses on the introduction of formal methods. Researchers should focus more on empirically showing that introducing a formal tool actually allows companies to reduce the cost of the testing phase, either by detecting errors beforehand or by facilitating the production of test cases. 

Our work shows that the usability of the tools is in general \textit{acceptable}. We also observe that there are conflicting reports regarding the ease of learning of tools that require advanced mathematical background. For example, Newcombe et al.~\cite{10.1145/2699417} reported that Amazon engineers with no prior knowledge of formal tools required only few weeks of training to utilize TLA+, and similar reports are available for the B method~\cite{Russo2013}. Therefore, circumstantial evidence seems to contradict common beliefs about the inherent complexity of learning formal methods, and we encourage more empirically grounded evaluations to answer the question: 
``Are formal methods truly difficult to learn?''. 

\textit{Implications for Railway Practitioners.} 
Several suggestions have been given to address the skill barriers, and these are mostly oriented to improve education of engineers~\cite{davis2013study,bjorner201440}. However, we argue that this may be a workable problem. To be effectively used, most formal tools currently require experts in formal methods and in the chosen tool~\cite{steffen2017physics,mazzanti2018towards}. Engineers need to understand the principles of a tool to interact with the  experts, but do not necessarily need to be experts themselves. Our usability study suggests that current tools may be sufficiently acceptable by engineers, when they see the tools used by another subject. Considering that also the maturity prejudice appears to be not well founded, we invite practitioners to overcome their notorious skepticism~\cite{woodcock2009formal,GM20}, and give formal tools a chance. Not in the least because Miller~\cite{Mil12} provides \textit{economical evidence} of the benefits of the application of formal methods and tools (including nuSMV, PVS, SAL, SCADE, and Simulink) to industrial problems in avionics. 

\section{Conclusion}
\label{sec:conclusion}
This paper presents a systematic evaluation of 13~formal tools for railway system design as well as a preliminary usability analysis of 7~of these formal tools. We show that the majority of the considered tools are mature and consolidated products, with a reasonably sufficient degree of industry readiness. Although many tools have limited GUIs, and require strong mathematical background, our usability study shows that, on average, the degree of usability of these 7~tools is between OK and Good, thus suggesting that usability may not be a strong barrier for formal tools' adoption. Main barriers are the limited support for development functionalities, such as traceability, and other process-integration features. We share our evaluation sheets~\cite{mazzanti2021data}, which include synthetic yet industry-relevant information about each single tool. 
 
Our work provides a contribution that follows the recommendations of Huisman et al.\cite{HGM20}, who ask different stakeholders---next to tool developers, researchers and practitioners, also policy makers and education staff---to join in an effort to reduce the gap between formal methods research and practice. Our study, and the shared data, can be useful to practitioners interested in introducing formal methods in their company, and to the research community of software engineers dealing with system development. Through our contribution, these two different profiles can have a clear and up-to-date opinion about the landscape of available tools, and their salient characteristics. Finally, tool developers can profit from our independent evaluation to identify the most suitable improvements for their tools.

\ifCLASSOPTIONcompsoc
  \section*{Acknowledgments}
\else
  \section*{Acknowledgment}
\fi

The authors would like to thank the ASTRail and 4SECURail projects, which received funding from the Shift2Rail Joint Undertaking under the European Union's Horizon 2020 research and innovation programme under grant agreements 777561 (ASTRail) and 881775 (4SECURail) in the context of the open call S2R-OC-IP2-01-2019, part of the \lq\lq Annual Work Plan and Budget 2019\rq\rq, of the programme H2020-S2RJU-2019. 
The content of this paper reflects only the authors' view and the Shift2Rail Joint Undertaking is not responsible for any use that may be made of the included information.

\ifCLASSOPTIONcaptionsoff
  \newpage
\fi



\bibliographystyle{IEEEtran}
\bibliography{bibliography}

\begin{thebibliography}{100}
\providecommand{\url}[1]{#1}
\csname url@samestyle\endcsname
\providecommand{\newblock}{\relax}
\providecommand{\bibinfo}[2]{#2}
\providecommand{\BIBentrySTDinterwordspacing}{\spaceskip=0pt\relax}
\providecommand{\BIBentryALTinterwordstretchfactor}{4}
\providecommand{\BIBentryALTinterwordspacing}{\spaceskip=\fontdimen2\font plus
\BIBentryALTinterwordstretchfactor\fontdimen3\font minus
  \fontdimen4\font\relax}
\providecommand{\BIBforeignlanguage}[2]{{%
\expandafter\ifx\csname l@#1\endcsname\relax
\typeout{** WARNING: IEEEtran.bst: No hyphenation pattern has been}%
\typeout{** loaded for the language `#1'. Using the pattern for}%
\typeout{** the default language instead.}%
\else
\language=\csname l@#1\endcsname
\fi
#2}}
\providecommand{\BIBdecl}{\relax}
\BIBdecl

\bibitem{ChiappiniCMRRSTV10}
A.~Chiappini, A.~Cimatti, L.~Macchi, O.~Rebollo, M.~Roveri, A.~Susi,
  S.~Tonetta, and B.~Vittorini, ``Formalization and validation of a subset of
  the {European} {Train} {Control} {System},'' in \emph{Proceedings of the 32nd
  {ACM/IEEE} International Conference on Software Engineering (ICSE'10)},
  vol.~2.\hskip 1em plus 0.5em minus 0.4em\relax ACM, 2010, pp. 109--118.

\bibitem{ferrari2013metro}
A.~Ferrari, D.~Grasso, G.~Magnani, A.~Fantechi, and M.~Tempestini, ``The
  {Metr{\^o}} {Rio} case study,'' \emph{Sci. Comput. Program.}, vol.~78, no.~7,
  pp. 828--842, 2013.

\bibitem{HaxthausenPK11}
A.~E. Haxthausen, J.~Peleska, and S.~Kinder, ``A formal approach for the
  construction and verification of railway control systems,'' \emph{Form. Asp.
  Comput.}, vol.~23, no.~2, pp. 191--219, 2011.

\bibitem{WR03}
K.~Winter and N.~J. Robinson, ``Modelling large railway interlockings and model
  checking small ones,'' in \emph{Proceedings of the 26th Australasian Computer
  Science Conference (ACSC'03)}, ser. CRPIT, M.~J. Oudshoorn, Ed.,
  vol.~16.\hskip 1em plus 0.5em minus 0.4em\relax Australian Computer Society,
  2003, pp. 309--316.

\bibitem{fantechi2013twenty}
A.~Fantechi, ``Twenty-five years of formal methods and railways: What next?''
  in \emph{Revised Selected Papers of the SEFM 2013 Collocated Workshops:
  BEAT2, WS-FMDS, FM-RAIL-Bok, MoKMaSD, and OpenCert}, ser. LNCS, S.~Counsell
  and M.~N{\'u}{\~{n}}ez, Eds., vol. 8368.\hskip 1em plus 0.5em minus
  0.4em\relax Springer, 2013, pp. 167--183.

\bibitem{FFG16}
A.~Fantechi, A.~Ferrari, and S.~Gnesi, ``Formal methods and safety
  certification: {C}hallenges in the railways domain,'' in \emph{Proceedings of
  the 7th International Symposium on Leveraging Applications of Formal Methods,
  Verification and Validation: Discussion, Dissemination, Applications
  (ISoLA'16)}, ser. LNCS, T.~Margaria and B.~Steffen, Eds., vol. 9953.\hskip
  1em plus 0.5em minus 0.4em\relax Springer, 2016, pp. 261--265.

\bibitem{wing1990specifier}
J.~M. Wing, ``A specifier's introduction to formal methods,'' \emph{IEEE
  Comp.}, vol.~23, no.~9, pp. 8--22, 1990.

\bibitem{GBP20}
H.~Garavel, M.~H. ter Beek, and J.~van~de Pol, ``{The 2020 Expert Survey on
  Formal Methods},'' in \emph{Proceedings of the 25th International Conference
  on Formal Methods for Industrial Critical Systems (FMICS'20)}, ser. LNCS,
  M.~H. ter Beek and D.~Ni{\v{c}}kovi{\'{c}}, Eds., vol. 12327.\hskip 1em plus
  0.5em minus 0.4em\relax Springer, 2020, pp. 3--69.

\bibitem{woodcock2009formal}
J.~Woodcock, P.~G. Larsen, J.~Bicarregui, and J.~Fitzgerald, ``Formal methods:
  Practice and experience,'' \emph{ACM Comput. Surv.}, vol.~41, no.~4, pp.
  19:1--19:36, 2009.

\bibitem{garavel2013}
H.~Garavel and S.~Graf, ``Formal methods for safe and secure computer
  systems,'' Bundesamt f\"{u}r Sicherheit in der Informationstechnik, Bonn,
  Germany, BSI Study 875, December 2013.

\bibitem{garavel2019reflections}
H.~Garavel and R.~Mateescu, ``Reflections on {Bernhard} {Steffen}’s physics
  of software tools,'' in \emph{Models, Mindsets, Meta: The What, the How, and
  the Why Not?}, ser. LNCS, T.~Margaria, S.~Graf, and K.~G. Larsen, Eds.\hskip
  1em plus 0.5em minus 0.4em\relax Springer, 2019, vol. 11200, pp. 186--207.

\bibitem{FBMBFGPT19}
A.~Ferrari, M.~H. ter Beek, F.~Mazzanti, D.~Basile, A.~Fantechi, S.~Gnesi,
  A.~Piattino, and D.~Trentini, ``Survey on formal methods and tools in
  railways: {T}he {ASTRail} approach,'' in \emph{Proceedings of the 3rd
  International Conference on Reliability, Safety, and Security of Railway
  Systems: Modelling, Analysis, Verification, and Certification (RSSRail'19)},
  ser. LNCS, S.~Collart-Dutilleul, T.~Lecomte, and A.~Romanovsky, Eds., vol.
  11495.\hskip 1em plus 0.5em minus 0.4em\relax Springer, 2019, pp. 226--241.

\bibitem{hartmanns2015quantitative}
A.~Hartmanns and H.~Hermanns, ``In the quantitative automata zoo,'' \emph{Sci.
  Comput. Program.}, vol. 112, pp. 3--23, 2015.

\bibitem{katoen2016probabilistic}
J.-P. Katoen, ``The probabilistic model checking landscape,'' in
  \emph{Proceedings of the 31st Annual {ACM/IEEE} Symposium on Logic in
  Computer Science (LICS'16)}.\hskip 1em plus 0.5em minus 0.4em\relax ACM,
  2016, pp. 31--45.

\bibitem{cenelec50128}
``{EN} 50128~-- {R}ailway applications~-- {C}ommunication, signalling and
  processing systems~-- {S}oftware for railway control and protection
  systems,'' CENELEC, June 2011.

\bibitem{davis2013study}
J.~A. Davis, M.~A. Clark, D.~D. Cofer, A.~Fifarek, J.~Hinchman, J.~A. Hoffman,
  B.~W. Hulbert, S.~P. Miller, and L.~G. Wagner, ``Study on the barriers to the
  industrial adoption of formal methods,'' in \emph{Proceedings of the 18th
  International Workshop on Formal Methods for Industrial Critical Systems
  (FMICS'13)}, ser. LNCS, C.~Pecheur and M.~Dierkes, Eds., vol. 8187.\hskip 1em
  plus 0.5em minus 0.4em\relax Springer, 2013, pp. 63--77.

\bibitem{nyberg2018formal}
M.~Nyberg, D.~Gurov, C.~Lidstr{\"{o}}m, A.~Rasmusson, and J.~Westman, ``Formal
  verification in automotive industry: {E}nablers and obstacles,'' in
  \emph{Proceedings of the 8th International Symposium on Leveraging
  Applications of Formal Methods, Verification and Validation: Industrial
  Practice (ISoLA'18)}, ser. LNCS, T.~Margaria and B.~Steffen, Eds., vol.
  11247.\hskip 1em plus 0.5em minus 0.4em\relax Springer, 2018, pp. 139--158.

\bibitem{FMBBF20}
A.~Ferrari, F.~Mazzanti, D.~Basile, M.~H. ter Beek, and A.~Fantechi,
  ``Comparing formal tools for system design: a judgment study,'' in
  \emph{Proceedings of the 42nd {ACM/IEEE} International Conference on Software
  Engineering (ICSE'20)}.\hskip 1em plus 0.5em minus 0.4em\relax ACM, 2020, pp.
  62--74.

\bibitem{schlick2018proposal}
R.~Schlick, M.~Felderer, I.~Majzik, R.~Nardone, A.~Raschke, C.~F. Snook, and
  V.~Vittorini, ``A proposal of an example and experiments repository to foster
  industrial adoption of formal methods,'' in \emph{Proceedings of the 8th
  International Symposium on Leveraging Applications of Formal Methods,
  Verification and Validation: Industrial Practice (ISoLA'18)}, ser. LNCS,
  T.~Margaria and B.~Steffen, Eds., vol. 11247.\hskip 1em plus 0.5em minus
  0.4em\relax Springer, 2018, pp. 249--272.

\bibitem{steffen2017physics}
B.~Steffen, ``The physics of software tools: {SWOT} analysis and vision,''
  \emph{Int. J. Softw. Tools Technol. Transf.}, vol.~19, no.~1, pp. 1--7, 2017.

\bibitem{abrial2007formal}
J.-R. Abrial, ``Formal methods: {T}heory becoming practice,'' \emph{J. Univers.
  Comput. Sci.}, vol.~13, no.~5, pp. 619--628, 2007.

\bibitem{leuschelFFP11}
M.~Leuschel, J.~Falampin, F.~Fritz, and D.~Plagge, ``{Automated property
  verification for large scale {B} models with {ProB}},'' \emph{Form. Asp.
  Comput.}, vol.~23, no.~6, pp. 683--709, 2011.

\bibitem{FFGM13}
A.~Ferrari, A.~Fantechi, S.~Gnesi, and G.~Magnani, ``Model-based development
  and formal methods in the railway industry,'' \emph{IEEE Softw.}, vol.~30,
  no.~3, pp. 28--34, 2013.

\bibitem{MNRST13}
F.~Moller, H.~N. Nguyen, M.~Roggenbach, S.~Schneider, and H.~Treharne,
  ``Defining and model checking abstractions of complex railway models using
  {CSP$\parallel$B},'' in \emph{Revised Selected Papers of the 8th
  International Haifa Verification Conference (HVC'12)}, ser. LNCS, A.~Biere,
  A.~Nahir, and T.~Vos, Eds., vol. 7857.\hskip 1em plus 0.5em minus 0.4em\relax
  Springer, 2013, pp. 193--208.

\bibitem{Gha14}
M.~Ghazel, ``Formalizing a subset of {ERTMS/ETCS} specifications for
  verification purposes,'' \emph{Transport. Res. C-Emer.}, vol.~42, pp. 60--75,
  2014.

\bibitem{JMNRST14}
P.~James, F.~Moller, H.~N. Nguyen, M.~Roggenbach, S.~Schneider, and
  H.~Treharne, ``Techniques for modelling and verifying railway
  interlockings,'' \emph{Int. J. Softw. Tools Technol. Transf.}, vol.~16,
  no.~6, pp. 685--711, 2014.

\bibitem{BosschaartQJG15}
M.~Bosschaart, E.~Quaglietta, B.~Janssen, and R.~M.~P. Goverde, ``Efficient
  formalization of railway interlocking data in {RailML},'' \emph{Inf. Syst.},
  vol.~49, pp. 126--141, 2015.

\bibitem{NGBPVMM16}
R.~Nardone, U.~Gentile, M.~Benerecetti, A.~Peron, V.~Vittorini, S.~Marrone, and
  N.~Mazzocca, ``Modeling railway control systems in {Promela},'' in
  \emph{Revised Selected Papers of the 4th International Workshop on Formal
  Techniques for Safety-Critical Systems (FTSCS'15)}, ser. CCIS, C.~Artho and
  P.~C. {\"{O}}lveczky, Eds., vol. 596.\hskip 1em plus 0.5em minus 0.4em\relax
  Springer, 2016, pp. 121--136.

\bibitem{BDG17}
D.~Basile, F.~{Di Giandomenico}, and S.~Gnesi, ``Statistical model checking of
  an energy-saving cyber-physical system in the railway domain,'' in
  \emph{Proceedings of the 32nd Symposium on Applied Computing (SAC'17)}.\hskip
  1em plus 0.5em minus 0.4em\relax ACM, 2017, pp. 1356--1363.

\bibitem{CLSQTL17}
Q.~Cappart, C.~Limbr{\'{e}}e, P.~Schaus, J.~Quilbeuf, L.~Traonouez, and
  A.~Legay, ``Verification of interlocking systems using statistical model
  checking,'' in \emph{Proceedings of the 18th International Symposium on High
  Assurance Systems Engineering (HASE'17)}.\hskip 1em plus 0.5em minus
  0.4em\relax IEEE, 2017, pp. 61--68.

\bibitem{Gha17}
M.~Ghazel, ``{A Control Scheme for Automatic Level Crossings Under the
  {ERTMS/ETCS} Level 2/3 Operation},'' \emph{IEEE Trans. Intell. Transp.
  Syst.}, vol.~18, no.~10, pp. 2667--2680, 2017.

\bibitem{LecomteDPM17}
T.~Lecomte, D.~D{\'{e}}harbe, {\'{E}}.~Prun, and E.~Mottin, ``Applying a formal
  method in industry: {A} 25-year trajectory,'' in \emph{Proceedings of the
  20th Brazilian Symposium on Formal Methods: Foundations and Applications
  (SBMF'17)}, ser. LNCS, S.~Cavalheiro and J.~Fiadeiro, Eds., vol. 10623.\hskip
  1em plus 0.5em minus 0.4em\relax Springer, 2017, pp. 70--87.

\bibitem{vu2017formal}
L.~H. Vu, A.~E. Haxthausen, and J.~Peleska, ``Formal modelling and verification
  of interlocking systems featuring sequential release,'' \emph{Sci. Comput.
  Program.}, vol. 133, pp. 91--115, 2017.

\bibitem{BLW18}
M.~Bartholomeus, B.~Luttik, and T.~Willemse, ``Modeling and analysing {ERTMS}
  hybrid level 3 with the {mCRL2} toolset,'' in \emph{Proceedings of the 23rd
  International Conference on Formal Methods for Industrial Critical Systems
  (FMICS'18)}, ser. LNCS, F.~Howar and J.~Barnat, Eds., vol. 11119.\hskip 1em
  plus 0.5em minus 0.4em\relax Springer, 2018, pp. 98--114.

\bibitem{BBC18}
D.~Basile, M.~H. ter Beek, and V.~Ciancia, ``Statistical model checking of a
  moving block railway signalling scenario with \textsc{Uppaal} {SMC}:
  {E}xperience and outlook,'' in \emph{Proceedings of the 8th International
  Symposium on Leveraging Applications of Formal Methods, Verification and
  Validation: Verification (ISoLA'18)}, ser. LNCS, T.~Margaria and B.~Steffen,
  Eds., vol. 11245.\hskip 1em plus 0.5em minus 0.4em\relax Springer, 2018, pp.
  372--391.

\bibitem{BergerJLRS18}
U.~Berger, P.~James, A.~Lawrence, M.~Roggenbach, and M.~Seisenberger,
  ``Verification of the {European} {Rail} {Traffic} {Management} {System} in
  {Real-Time} {Maude},'' \emph{Sci. Comput. Program.}, vol. 154, pp. 61--88,
  2018.

\bibitem{IliasovTLR18}
A.~Iliasov, D.~Taylor, L.~Laibinis, and A.~B. Romanovsky, ``Formal verification
  of signalling programs with {SafeCap},'' in \emph{Proceedings of the 37th
  International Conference on Computer Safety, Reliability, and Security
  (SAFECOMP 2018)}, ser. LNCS, B.~Gallina, A.~Skavhaug, and F.~Bitsch, Eds.,
  vol. 11093.\hskip 1em plus 0.5em minus 0.4em\relax Springer, 2018, pp.
  91--106.

\bibitem{MF18}
F.~Mazzanti and A.~Ferrari, ``{Ten Diverse Formal Models for a {CBTC} Automatic
  Train Supervision System},'' in \emph{Proceedings of the 3rd Workshop on
  Models for Formal Analysis of Real Systems and the 6th International Workshop
  on Verification and Program Transformation (MARS/VPT'18)}, ser. EPTCS, J.~P.
  Gallagher, R.~van Glabbeek, and W.~Serwe, Eds., vol. 268, 2018, pp. 104--149.

\bibitem{Thai18}
S.~Vanit-Anunchai, ``Modelling and simulating a {Thai} railway signalling
  system using {Coloured} {Petri} {Nets},'' \emph{Int. J. Softw. Tools Technol.
  Transf.}, vol.~20, no.~3, pp. 243--262, 2018.

\bibitem{BBFL19}
D.~Basile, M.~H. ter Beek, A.~Ferrari, and A.~Legay, ``Modelling and analysing
  {ERTMS} {L3} moving block railway signalling with {Simulink} and
  \textsc{Uppaal} {SMC},'' in \emph{Proceedings of the 24th International
  Conference on Formal Methods for Industrial Critical Systems (FMICS'19)},
  ser. LNCS, K.~G. Larsen and T.~Willemse, Eds., vol. 11687.\hskip 1em plus
  0.5em minus 0.4em\relax Springer, 2019, pp. 1--21.

\bibitem{Abr20}
J.~Abrial, ``The {ABZ-2018} case study with {Event-B},'' \emph{Int. J. Softw.
  Tools Technol. Transf.}, vol.~22, no.~3, pp. 257--264, 2020.

\bibitem{AKJ20}
P.~Arcaini, J.~Kofro{\v{n}}, and P.~Je{\v{z}}ek, ``Validation of the hybrid
  {ERTMS/ETCS} level 3 using {\sc spin},'' \emph{Int. J. Softw. Tools Technol.
  Transf.}, vol.~22, no.~3, pp. 265--279, 2020.

\bibitem{CM20}
A.~Cunha and N.~Macedo, ``Validating the hybrid {ERTMS/ETCS} level 3 concept
  with {Electrum},'' \emph{Int. J. Softw. Tools Technol. Transf.}, vol.~22,
  no.~3, pp. 281--296, 2020.

\bibitem{DDPS20}
D.~Dghaym, M.~Dalvandi, M.~Poppleton, and C.~Snook, ``Formalising the hybrid
  {ERTMS} level 3 specification in {iUML-B} and {Event-B},'' \emph{Int. J.
  Softw. Tools Technol. Transf.}, vol.~22, no.~3, pp. 297--313, 2020.

\bibitem{HLKKNNSS20}
D.~Hansen, M.~Leuschel, P.~K{\"{o}}rner, S.~Krings, T.~Naulin, N.~Nayeri,
  D.~Schneider, and F.~Skowron, ``Validation and real-life demonstration of
  {ETCS} hybrid level 3 principles using a formal {B} model,'' \emph{Int. J.
  Softw. Tools Technol. Transf.}, vol.~22, no.~3, pp. 315--332, 2020.

\bibitem{MFTL20}
A.~Mammar, M.~Frappier, S.~J. {Tueno Fotso}, and R.~Laleau, ``A formal
  refinement-based analysis of the hybrid {ERTMS/ETCS} level 3 standard,''
  \emph{Int. J. Softw. Tools Technol. Transf.}, vol.~22, no.~3, pp. 333--347,
  2020.

\bibitem{TFLM20}
S.~J. {Tueno Fotso}, M.~Frappier, R.~Laleau, and A.~Mammar, ``Modeling the
  hybrid {ERTMS/ETCS} level 3 standard using a formal requirements engineering
  approach,'' \emph{Int. J. Softw. Tools Technol. Transf.}, vol.~22, no.~3, pp.
  349--363, 2020.

\bibitem{BBL20}
D.~Basile, M.~H. ter Beek, and A.~Legay, ``Strategy synthesis for autonomous
  driving in a moving block railway system with \textsc{Uppaal Stratego},'' in
  \emph{Proceedings of the 40th {IFIP} {WG} 6.1 International Conference on
  Formal Techniques for Distributed Objects, Components, and Systems
  (FORTE'20)}, ser. LNCS, A.~Gotsman and A.~Sokolova, Eds., vol. 12136.\hskip
  1em plus 0.5em minus 0.4em\relax Springer, 2020, pp. 3--21.

\bibitem{BBF0GMMPT20}
D.~Basile, M.~H. ter Beek, A.~Fantechi, A.~Ferrari, S.~Gnesi, L.~Masullo,
  F.~Mazzanti, A.~Piattino, and D.~Trentini, ``Designing a demonstrator of
  formal methods for railways infrastructure managers,'' in \emph{Proceedings
  of the 9th International Symposium on Leveraging Applications of Formal
  Methods: Applications (ISoLA'20)}, ser. LNCS, T.~Margaria and B.~Steffen,
  Eds., vol. 12478.\hskip 1em plus 0.5em minus 0.4em\relax Springer, 2020, pp.
  467--485.

\bibitem{mazzanti2018towards}
F.~Mazzanti, A.~Ferrari, and G.~O. Spagnolo, ``{Towards formal methods
  diversity in railways: an experience report with seven frameworks},''
  \emph{Int. J. Softw. Tools Technol. Transf.}, vol.~20, no.~3, pp. 263--288,
  2018.

\bibitem{haxthausen16comparing}
A.~E. Haxthausen, H.~N. Nguyen, and M.~Roggenbach, ``Comparing formal
  verification approaches of interlocking systems,'' in \emph{Proceedings of
  the 1st International Conference on Reliability, Safety, and Security of
  Railway Systems: Modelling, Analysis, Verification, and Certification
  (RSSRail'16)}, ser. LNCS, T.~Lecomte, R.~Pinger, and A.~B. Romanovsky, Eds.,
  vol. 9707.\hskip 1em plus 0.5em minus 0.4em\relax Springer, 2016, pp.
  160--177.

\bibitem{BBGFGS20}
D.~Basile, M.~H. ter Beek, F.~D. Giandomenico, A.~Fantechi, S.~Gnesi, and G.~O.
  Spagnolo, ``30 years of simulation-based quantitative analysis tools: {A}
  comparison experiment between {M{\"{o}}bius} and {Uppaal} {SMC},'' in
  \emph{Proceedings of the 9th International Symposium on Leveraging
  Applications of Formal Methods: Verification Principles (ISoLA'20)}, ser.
  LNCS, T.~Margaria and B.~Steffen, Eds., vol. 12476.\hskip 1em plus 0.5em
  minus 0.4em\relax Springer, 2020, pp. 368--384.

\bibitem{kitchenham1997desmet}
B.~Kitchenham, S.~Linkman, and D.~Law, ``{{DESMET}: a methodology for
  evaluating software engineering methods and tools},'' \emph{Comput. Control.
  Eng. J.}, vol.~8, no.~3, pp. 120--126, 1997.

\bibitem{BBFGMPTF18}
D.~Basile, M.~H. ter Beek, A.~Fantechi, S.~Gnesi, F.~Mazzanti, A.~Piattino,
  D.~Trentini, and A.~Ferrari, ``On the industrial uptake of formal methods in
  the railway domain: {A} survey with stakeholders,'' in \emph{Proceedings of
  the 14th International Conference on Integrated Formal Methods (iFM'18)},
  ser. LNCS, C.~A. Furia and K.~Winter, Eds., vol. 11023.\hskip 1em plus 0.5em
  minus 0.4em\relax Springer, 2018, pp. 20--29.

\bibitem{BBFFGLM19}
M.~H. ter Beek, A.~Bor{\"{a}}lv, A.~Fantechi, A.~Ferrari, S.~Gnesi,
  C.~L{\"{o}}fving, and F.~Mazzanti, ``Adopting formal methods in an industrial
  setting: {T}he railways case,'' in \emph{Proceedings of the 3rd World
  Congress on Formal Methods (FM'19)}, ser. LNCS, M.~H. ter Beek, A.~McIver,
  and J.~N. Oliveira, Eds., vol. 11800.\hskip 1em plus 0.5em minus 0.4em\relax
  Springer, 2019, pp. 762--772.

\bibitem{mazzanti2021data}
\BIBentryALTinterwordspacing
F.~Mazzanti, A.~Ferrari, D.~Basile, and M.~ter Beek, ``{Systematic Evaluation
  and Usability Analysis of Formal Tools for Railway System Design - Technical
  Annexes},'' Zenodo, April 2021. [Online]. Available:
  \url{https://doi.org/10.5281/zenodo.4675769}
\BIBentrySTDinterwordspacing

\bibitem{BGK18}
M.~H. ter Beek, S.~Gnesi, and A.~Knapp, ``Formal methods for transport
  systems,'' \emph{Int. J. Softw. Tools Technol. Transf.}, vol.~20, no.~3, pp.
  237--241, 2018.

\bibitem{Duf91}
D.~Duffy, \emph{Principles of Automated Theorem Proving}.\hskip 1em plus 0.5em
  minus 0.4em\relax John Wiley \& Sons, 1991.

\bibitem{RV01}
J.~A. Robinson and A.~Voronkov, Eds., \emph{Handbook of Automated
  Reasoning}.\hskip 1em plus 0.5em minus 0.4em\relax Elsevier, 2001.

\bibitem{bertot2013interactive}
Y.~Bertot and P.~Cast{\'e}ran, \emph{Interactive Theorem Proving and Program
  Development}.\hskip 1em plus 0.5em minus 0.4em\relax Springer, 2004.

\bibitem{CGP99}
E.~M. Clarke, O.~Grumberg, and D.~A. Peled, \emph{Model Checking}.\hskip 1em
  plus 0.5em minus 0.4em\relax MIT Press, 1999.

\bibitem{BK08}
C.~Baier and J.-P. Katoen, \emph{Principles of Model Checking}.\hskip 1em plus
  0.5em minus 0.4em\relax MIT Press, 2008.

\bibitem{CHVB18}
E.~M. Clarke, T.~A. Henzinger, H.~Veith, and R.~Bloem, Eds., \emph{Handbook of
  Model Checking}.\hskip 1em plus 0.5em minus 0.4em\relax Springer, 2018.

\bibitem{gibson2014fdr3}
T.~Gibson{-}Robinson, P.~J. Armstrong, A.~Boulgakov, and A.~W. Roscoe,
  ``{FDR3}: a parallel refinement checker for {CSP},'' \emph{Int. J. Softw.
  Tools Technol. Transf.}, vol.~18, no.~2, pp. 149--167, 2016.

\bibitem{atelierb}
\emph{Atelier~B User Manual Version 4.0}, ClearSy, 2002.

\bibitem{Ber08}
Y.~Bertot, ``A short presentation of {Coq},'' in \emph{Proceedings of the 21st
  International Conference on Theorem Proving in Higher Order Logics
  (TPHOLs'08)}, ser. LNCS, O.~A. Mohamed, C.~A. Mu{\~{n}}oz, and S.~Tahar,
  Eds., vol. 5170.\hskip 1em plus 0.5em minus 0.4em\relax Springer, 2008, pp.
  12--16.

\bibitem{paulson1994isabelle}
L.~C. Paulson, \emph{Isabelle: A Generic Theorem Prover}, ser. LNCS.\hskip 1em
  plus 0.5em minus 0.4em\relax Springer, 1994, vol. 828.

\bibitem{OS08}
S.~Owre and N.~Shankar, ``A brief overview of {PVS},'' in \emph{Proceedings of
  the 21st International Conference on Theorem Proving in Higher Order Logics
  (TPHOLs'08)}, ser. LNCS, O.~A. Mohamed, C.~A. Mu{\~{n}}oz, and S.~Tahar,
  Eds., vol. 5170.\hskip 1em plus 0.5em minus 0.4em\relax Springer, 2008, pp.
  22--27.

\bibitem{holzmann2004spin}
G.~J. Holzmann, \emph{The Spin Model Checker: Primer and Reference
  Manual}.\hskip 1em plus 0.5em minus 0.4em\relax Addison-Wesley, 2003.

\bibitem{TERBEEK2011119}
M.~H. ter Beek, A.~Fantechi, S.~Gnesi, and F.~Mazzanti, ``A state/event-based
  model-checking approach for the analysis of abstract system properties,''
  \emph{Sci. Comput. Program.}, vol.~76, no.~2, pp. 119--135, 2011.

\bibitem{McM93}
K.~L. McMillan, \emph{Symbolic Model Checking}.\hskip 1em plus 0.5em minus
  0.4em\relax Kluwer, 1993.

\bibitem{CCCGR00}
A.~Cimatti, E.~M. Clarke, F.~Giunchiglia, and M.~Roveri, ``{\sc NuSMV}: A new
  symbolic model checker,'' \emph{Int. J. Softw. Tools Technol. Transf.},
  vol.~2, no.~4, pp. 410--425, 2000.

\bibitem{CCDGMMMRT14}
R.~Cavada, A.~Cimatti, M.~Dorigatti, A.~Griggio, A.~Mariotti, A.~Micheli,
  S.~Mover, M.~Roveri, and S.~Tonetta, ``The {nuXmv} symbolic model checker,''
  in \emph{Proceedings of the 26th International Conference on Computer Aided
  Verification (CAV'14)}, ser. LNCS, A.~Biere and R.~Bloem, Eds., vol.
  8559.\hskip 1em plus 0.5em minus 0.4em\relax Springer, 2014, pp. 334--342.

\bibitem{KNP11}
M.~Kwiatkowska, G.~Norman, and D.~Parker, ``{PRISM} 4.0: Verification of
  probabilistic real-time systems,'' in \emph{Proceedings of the 23rd
  International Conference on Computer Aided Verification (CAV'11)}, ser. LNCS,
  G.~Gopalakrishnan and S.~Qadeer, Eds., vol. 6806.\hskip 1em plus 0.5em minus
  0.4em\relax Springer, 2011, pp. 585--591.

\bibitem{AP18}
G.~Agha and K.~Palmskog, ``A survey of statistical model checking,'' \emph{ACM
  Trans. Model. Comput. Simul.}, vol.~28, no.~1, pp. 6:1--6:39, 2018.

\bibitem{LLTYSG19}
A.~Legay, A.~Lukina, L.~Traonouez, J.~Yang, S.~A. Smolka, and R.~Grosu,
  ``Statistical model checking,'' in \emph{Computing and Software Science:
  State of the Art and Perspectives}, ser. LNCS, B.~Steffen and G.~J.
  Woeginger, Eds.\hskip 1em plus 0.5em minus 0.4em\relax Springer, 2019, vol.
  10000, pp. 478--504.

\bibitem{David2015}
A.~David, K.~G. Larsen, A.~Legay, M.~Miku{\v{c}}ionis, and D.~B. Poulsen,
  ``\textsc{Uppaal} {SMC} tutorial,'' \emph{Int. J. Softw. Tools Technol.
  Transf.}, vol.~17, no.~4, pp. 397--415, 2015.

\bibitem{CADP}
H.~Garavel, F.~Lang, R.~Mateescu, and W.~Serwe, ``{CADP} 2011: a toolbox for
  the construction and analysis of distributed processes,'' \emph{Int. J.
  Softw. Tools Technol. Transf.}, vol.~15, no.~2, pp. 89--107, 2013.

\bibitem{mCRL2}
J.~F. Groote and M.~R. Mousavi, \emph{Modeling and Analysis of Communicating
  Systems}.\hskip 1em plus 0.5em minus 0.4em\relax MIT Press, 2014.

\bibitem{10.5555/579617}
L.~Lamport, \emph{Specifying Systems: The {TLA+} Language and Tools for
  Hardware and Software Engineers}.\hskip 1em plus 0.5em minus 0.4em\relax
  Addison-Wesley, 2002.

\bibitem{10.1007/978-3-540-27813-9_45}
L.~M. de~Moura, S.~Owre, H.~Rue{\ss}, J.~M. Rushby, N.~Shankar, M.~Sorea, and
  A.~Tiwari, ``{SAL} 2,'' in \emph{Proceedings of the 16th International
  Conference on Computer Aided Verification (CAV'04)}, ser. LNCS, R.~Alur and
  D.~A. Peled, Eds., vol. 3114.\hskip 1em plus 0.5em minus 0.4em\relax
  Springer, 2004, pp. 496--500.

\bibitem{JKW07}
K.~Jensen, L.~M. Kristensen, and L.~Wells, ``{Coloured} {Petri} {Nets} and
  {CPN} {Tools} for modelling and validation of concurrent systems,''
  \emph{Int. J. Softw. Tools Technol. Transf.}, vol.~9, no. 3-4, pp. 213--254,
  2007.

\bibitem{modelica}
P.~Fritzson, \emph{Principles of Object Oriented Modeling and Simulation with
  {Modelica} 3.3: {A} Cyber‐Physical Approach}.\hskip 1em plus 0.5em minus
  0.4em\relax IEEE/Wiley, 2014.

\bibitem{10.1007/978-1-4020-6254-4_2}
G.~Berry, ``{SCADE}: {S}ynchronous design and validation of embedded control
  software,'' in \emph{Proceedings of the General Motors R\&D Workshop},
  S.~Ramesh and P.~Sampath, Eds.\hskip 1em plus 0.5em minus 0.4em\relax
  Springer, 2007, pp. 19--33.

\bibitem{dabney2004mastering}
J.~B. Dabney and T.~L. Harman, \emph{Mastering Simulink}.\hskip 1em plus 0.5em
  minus 0.4em\relax Pearson, 2003.

\bibitem{GM20}
M.~Gleirscher and D.~Marmsoler, ``Formal methods in dependable systems
  engineering: {A} survey of professionals from {Europe} and {North America},''
  \emph{Empir. Softw. Eng.}, vol.~25, no.~6, pp. 4473--4546, 2020.

\bibitem{GCR94}
S.~L. Gerhart, D.~Craigen, and T.~Ralston, ``Experience with formal methods in
  critical systems,'' \emph{IEEE Softw.}, vol.~11, no.~1, pp. 21--28, January
  1994.

\bibitem{ACJMPSO96}
M.~A. Ardis, J.~A. Chaves, L.~J. Jagadeesan, P.~Mataga, C.~Puchol, M.~G.
  Staskauskas, and J.~V. Olnhausen, ``A framework for evaluating specification
  methods for reactive systems: Experience report,'' \emph{IEEE Trans. Softw.
  Eng.}, vol.~22, no.~6, pp. 378--389, 1996.

\bibitem{ABDS18}
A.~Abbassi, A.~Bandali, N.~A. Day, and J.~Serna, ``A comparison of the
  declarative modelling languages {B}, {Dash}, and {TLA\textsuperscript{+}},''
  in \emph{Proceedings of the 8th {IEEE} International Model-Driven
  Requirements Engineering Workshop (MoDRE@RE'18)}.\hskip 1em plus 0.5em minus
  0.4em\relax IEEE, 2018, pp. 11--20.

\bibitem{bartocci2019toolympics}
E.~Bartocci, D.~Beyer, P.~E. Black, G.~Fedyukovich, H.~Garavel, A.~Hartmanns,
  M.~Huisman, F.~Kordon, J.~Nagele, M.~Sighireanu, B.~Steffen, M.~Suda,
  G.~Sutcliffe, T.~Weber, and A.~Yamada, ``{TOOLympics} 2019: {A}n overview of
  competitions in formal methods,'' in \emph{Proceedings of the 25th
  International Conference on Tools and Algorithms for the Construction and
  Analysis of Systems (TACAS'19)}, ser. LNCS, D.~Beyer, M.~Huisman, F.~Kordon,
  and B.~Steffen, Eds., vol. 11429.\hskip 1em plus 0.5em minus 0.4em\relax
  Springer, 2019, pp. 3--24.

\bibitem{steffen1997electronic}
B.~Steffen, T.~Margaria, and V.~Braun, ``The {Electronic} {Tool} {Integration}
  platform: concepts and design,'' \emph{Int. J. Softw. Tools Technol.
  Transf.}, vol.~1, no. 1-2, pp. 9--30, 1997.

\bibitem{bjorner2003new}
D.~Bj{\o}rner, ``New results and trends in formal techniques \& tools for the
  development of software for transportation systems: {A} review,'' in
  \emph{Proceedings of the 4th Symposium on Formal Methods for Railway
  Operation and Control Systems (FORMS'03)}, G.~Tarnai and E.~Schnieder,
  Eds.\hskip 1em plus 0.5em minus 0.4em\relax L'Harmattan, 2003.

\bibitem{FWM13}
A.~Fantechi, W.~Fokkink, and A.~Morzenti, ``Some trends in formal methods
  applications to railway signaling,'' in \emph{FMICS: A Survey of
  Applications}, S.~Gnesi and T.~Margaria, Eds.\hskip 1em plus 0.5em minus
  0.4em\relax IEEE/Wiley, 2013, ch.~4, pp. 61--84.

\bibitem{flammini2012railway}
F.~Flammini, Ed., \emph{Railway Safety, Reliability, and Security: Technologies
  and Systems Engineering}.\hskip 1em plus 0.5em minus 0.4em\relax IGI Global,
  2012.

\bibitem{boulanger2014formal}
J.-L. Boulanger, Ed., \emph{Formal Methods Applied to Industrial Complex
  Systems: {I}mplementation of the {B} Method}.\hskip 1em plus 0.5em minus
  0.4em\relax John Wiley \& Sons, 2014.

\bibitem{hoang2018hybrid}
T.~S. Hoang, M.~J. Butler, and K.~Reichl, ``The hybrid {ERTMS/ETCS} level 3
  case study,'' in \emph{Proceedings of the 6th International Conference on
  Abstract State Machines, Alloy, {B}, TLA, VDM, and {Z} (ABZ'18)}, ser. LNCS,
  M.~J. Butler, A.~Raschke, T.~S. Hoang, and K.~Reichl, Eds., vol. 10817.\hskip
  1em plus 0.5em minus 0.4em\relax Springer, 2018, pp. 251--261.

\bibitem{BHRR20}
M.~Butler, T.~S. Hoang, A.~Raschke, and K.~Reichl, ``Introduction to the
  special section on the {ABZ 2018} case study: Hybrid {ERTMS/ETCS} level 3,''
  \emph{Int. J. Softw. Tools Technol. Transf.}, vol.~22, no.~3, pp. 249--255,
  2020.

\bibitem{newcombe2014amazon}
C.~Newcombe, ``Why {Amazon} chose {TLA\textsuperscript{+}},'' in
  \emph{Proceedings of the 4th International Conference on Abstract State
  Machines, Alloy, {B}, TLA, VDM, and {Z} (ABZ'14)}, ser. LNCS, Y.~A. Ameur and
  K.~Schewe, Eds., vol. 8477.\hskip 1em plus 0.5em minus 0.4em\relax Springer,
  2014, pp. 25--39.

\bibitem{K04}
B.~Kitchenham, ``Procedures for performing systematic reviews,'' Keele
  University, Tech. Rep. TR/SE-0401, July 2004, and {N}ational {ICT}
  {A}ustralia, Tech. Rep. 0400011T.1.

\bibitem{scupin1997kj}
R.~Scupin, ``The {KJ} method: {A} technique for analyzing data derived from
  {Japanese} ethnology,'' \emph{Hum. Organ.}, vol.~56, no.~2, pp. 233--237,
  1997.

\bibitem{brooke1996sus}
J.~Brooke, ``{SUS}: A `quick and dirty' usability scale,'' in \emph{Usability
  Evaluation in Industry}, P.~W. Jordan, B.~Thomas, B.~A. Weerdmeester, and
  I.~L. McClelland, Eds.\hskip 1em plus 0.5em minus 0.4em\relax CRC press,
  1996, ch.~21, pp. 189--194.

\bibitem{Bro13}
------, ``{SUS}: A retrospective,'' \emph{J. Usability Stud.}, vol.~8, no.~2,
  pp. 29--40, 2013.

\bibitem{pourali2018empirical}
P.~Pourali and J.~M. Atlee, ``An empirical investigation to understand the
  difficulties and challenges of software modellers when using modelling
  tools,'' in \emph{Proceedings of the 21th {ACM/IEEE} International Conference
  on Model Driven Engineering Languages and Systems (MODELS'18)}.\hskip 1em
  plus 0.5em minus 0.4em\relax ACM, 2018, pp. 224--234.

\bibitem{bangor2008empirical}
A.~Bangor, P.~T. Kortum, and J.~T. Miller, ``An empirical evaluation of the
  system usability scale,'' \emph{Int. J. Hum. Comput. Interact.}, vol.~24,
  no.~6, pp. 574--594, 2008.

\bibitem{hwang2010number}
W.~Hwang and G.~Salvendy, ``Number of people required for usability evaluation:
  {T}he 10$\pm$2 rule,'' \emph{Commun. ACM}, vol.~53, no.~5, pp. 130--133,
  2010.

\bibitem{FHAB17}
N.~Furness, H.~van Houten, L.~Arenas, and M.~Bartholomeus, ``{ERTMS} level 3:
  the game-changer,'' \emph{IRSE News}, vol. 232, pp. 2--9, April 2017.

\bibitem{NEAR2050}
P.~Stroh, P.~Melia, and S.~Bilgici, ``Deliverable {D2.1}: {D}etermine the
  long-term changes in future needs,'' NEAR2050, January 2018.

\bibitem{standard2019ergonomics}
``{ISO} 9241-11:2018~-- {E}rgonomics of human-system interaction~-- {P}art~11:
  {U}sability: {D}efinitions and concepts,'' {ISO/TC~159/SC~4}, March 2018.

\bibitem{rierson2017developing}
L.~Rierson, \emph{Developing Safety-Critical Software: A Practical Guide for
  Aviation Software and {DO-178C} Compliance}.\hskip 1em plus 0.5em minus
  0.4em\relax CRC Press, 2017.

\bibitem{ameur2010toward}
Y.~A. Ameur, F.~Boniol, and V.~Wiels, ``Toward a wider use of formal methods
  for aerospace systems design and verification,'' \emph{Int. J. Softw. Tools
  Technol. Transf.}, vol.~12, no.~1, pp. 1--7, 2010.

\bibitem{atlee2013recommendations}
J.~M. Atlee, S.~Beidu, N.~A. Day, F.~Faghih, and P.~Shaker, ``Recommendations
  for improving the usability of formal methods for product lines,'' in
  \emph{Proceedings of the 1st {FME} Workshop on Formal Methods in Software
  Engineering (FormaliSE'13)}.\hskip 1em plus 0.5em minus 0.4em\relax IEEE,
  2013, pp. 43--49.

\bibitem{lamsweerde2000formal}
A.~van Lamsweerde, ``Formal specification: a roadmap,'' in \emph{Proceedings of
  the Future of Software Engineering Track of the 22nd International Conference
  on Software Engineering (ICSE'00)}.\hskip 1em plus 0.5em minus 0.4em\relax
  ACM, 2000, pp. 147--159.

\bibitem{ottensooser2012making}
A.~Ottensooser, A.~D. Fekete, H.~A. Reijers, J.~Mendling, and C.~Menictas,
  ``Making sense of business process descriptions: {A}n experimental comparison
  of graphical and textual notations,'' \emph{J. Syst. Softw.}, vol.~85, no.~3,
  pp. 596--606, 2012.

\bibitem{moscato2017automatic}
M.~M. Moscato, L.~Titolo, A.~Dutle, and C.~A. Mu{\~{n}}oz, ``Automatic
  estimation of verified floating-point round-off errors via static analysis,''
  in \emph{Proceedings of the 36th International Conference on Computer Safety,
  Reliability, and Security (SAFECOMP'17)}, ser. LNCS, S.~Tonetta,
  E.~Schoitsch, and F.~Bitsch, Eds., vol. 10488.\hskip 1em plus 0.5em minus
  0.4em\relax Springer, 2017, pp. 213--229.

\bibitem{gleirscher2019new}
M.~Gleirscher, S.~Foster, and J.~Woodcock, ``New opportunities for integrated
  formal methods,'' \emph{ACM Comput. Surv.}, vol.~52, no.~6, pp.
  117:1--117:36, 2020.

\bibitem{bjorner201440}
D.~Bj{\o}rner and K.~Havelund, ``40 years of formal methods: {S}ome obstacles
  and some possibilities?'' in \emph{Proceedings of the 19th International
  Symposium on Formal Methods (FM'14)}, ser. LNCS, C.~Jones, P.~Pihlajasaari,
  and J.~Sun, Eds., vol. 8442.\hskip 1em plus 0.5em minus 0.4em\relax Springer,
  2014, pp. 42--61.

\bibitem{10.1145/2699417}
C.~Newcombe, T.~Rath, F.~Zhang, B.~Munteanu, M.~Brooker, and M.~Deardeuff,
  ``How {Amazon} {Web} {Services} uses formal methods,'' \emph{Commun. ACM},
  vol.~58, no.~4, pp. 66--73, 2015.

\bibitem{Russo2013}
A.~G. Russo~Jr., ``Formal methods as an improvement tool,'' in \emph{Industrial
  Deployment of System Engineering Methods}, A.~B. Romanovsky and M.~Thomas,
  Eds.\hskip 1em plus 0.5em minus 0.4em\relax Springer, 2013, pp. 81--95.

\bibitem{Mil12}
S.~P. Miller, ``Lessons from twenty years of industrial formal methods,'' in
  \emph{Proceedings of the 20th High Confidence Software and Systems Conference
  (HCSS'12)}, 2012.

\bibitem{HGM20}
M.~Huisman, D.~Gurov, and A.~Malkis, ``Formal methods: From academia to
  industrial practice. {A} travel guide,'' arXiv:2002.07279 {\bf [cs.SE]},
  February 2020.

\end{thebibliography}
%
%
%
\newpage
%

\begin{IEEEbiography}[{\includegraphics[width=1in,height=1.25in,clip,keepaspectratio]{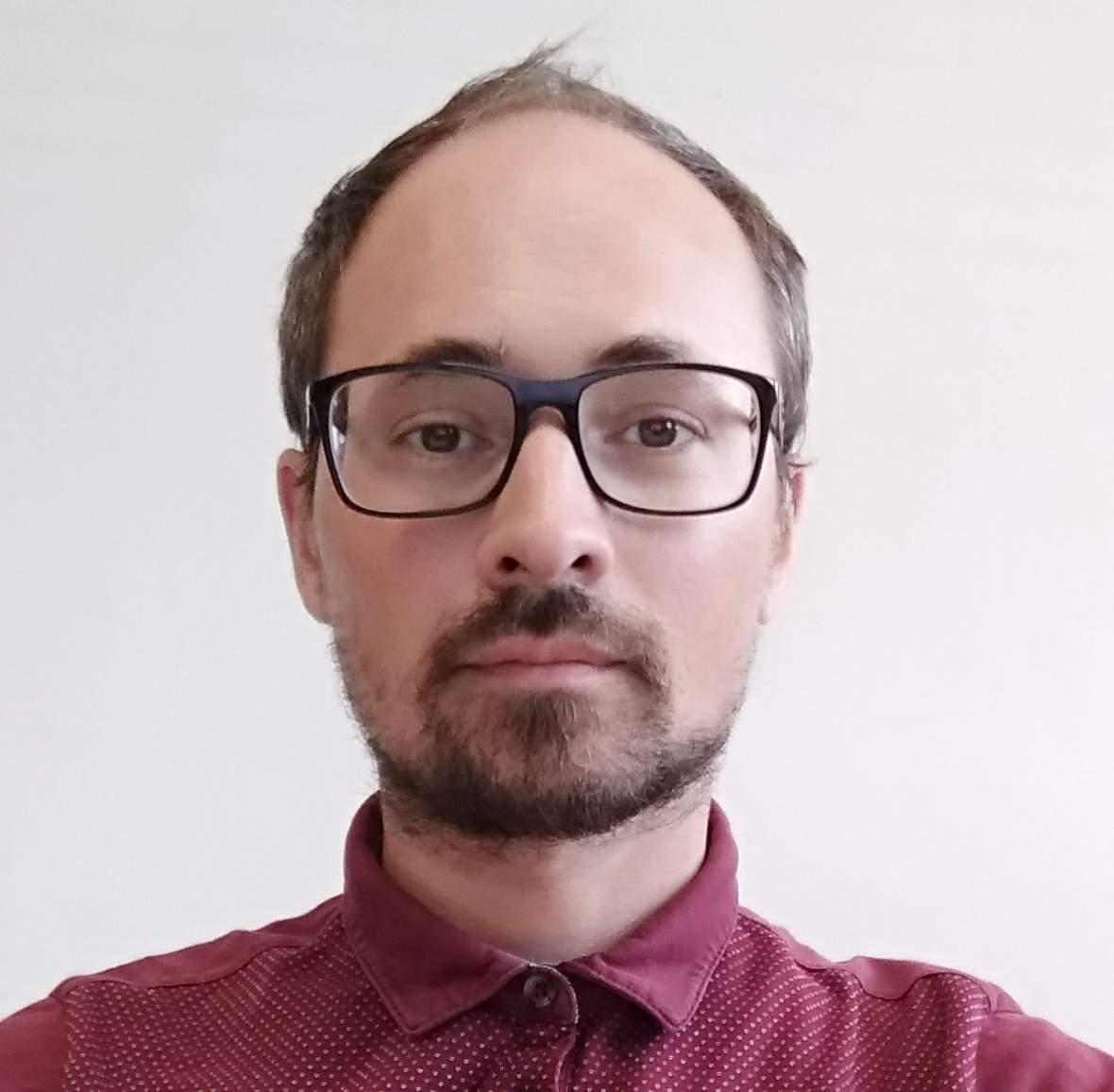}}]{Alessio Ferrari} is a research scientist at ISTI-CNR (Pisa, Italy) and member of the Formal Methods and Tools lab. His main research interests are requirements engineering,  empirical software engineering, and applications of formal methods. In these fields, he has published over 80 peer-reviewed papers in conferences and journals. He has been working as system engineer at General Electric Transportation Systems s.p.a., a world leading railway signalling manufacturer, where he used formal methods and code generation for the development of railway systems. He has been workpackage leader of the ASTRail project, and he is currently involved in the 4SECURail project, funded under the Shift2Rail initiative. 

\end{IEEEbiography}

\begin{IEEEbiography}[{\includegraphics[width=1in,height=1.25in,clip,keepaspectratio]{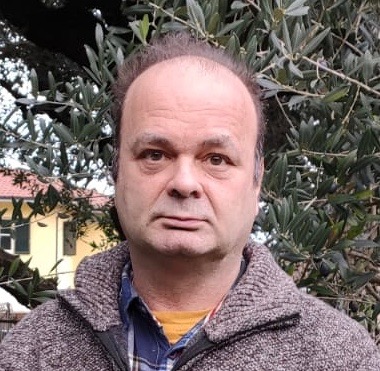}}]{Franco Mazzanti}
Franco Mazzanti is a senior researcher at ISTI--CNR (Pisa, Italy) and member of the Formal Methods and Tools lab. He is the main designer and author of the KandISTI formal verification framework, which includes the explicit, on-the-fly model-checking tools CMC, FMC, UMC and VMC. He has authored more than 40~papers in the field of formal methods. His research focuses on i)~the design and support of state- and event-based branching-time temporal logics for the specification and evaluation of system requirements, and ii)~the exploitation of formal methods and tools diversity for the analysis of concurrent, asynchronous systems. He has applied his research in many EU projects, among which ASTRail and 4SECURail.
\end{IEEEbiography}


\begin{IEEEbiography}[{\includegraphics[width=1in,height=1.25in,clip,keepaspectratio]{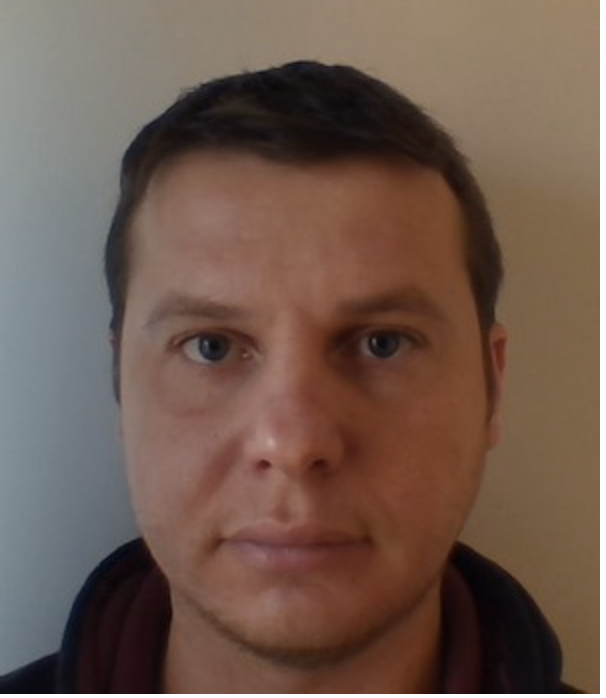}}]{Davide Basile}
Davide Basile is a researcher at ISTI–CNR (Pisa, Italy) and a member of 
the Formal Methods and Tools lab.
He has authored about 40 peer-reviewed papers in conferences and 
journals in the field of formal methods, software engineering and 
dependable computing.
His research focuses on developing both novel formalisms for emerging 
computational paradigms and
supporting tools, exploring formal verification techniques and applying 
state-of-the-art
formal methods to the design of real-world systems and emerging 
technologies in the railway domain.
He participated in several projects in collaboration with railway 
companies, including SISTER, with the participation
of Thales s.p.a., and ASTRail. He is currently involved in the 4SECURail 
project, funded under the Shift2Rail initiative.

\end{IEEEbiography}

\begin{IEEEbiography}[{\includegraphics[width=1in,height=1.25in,clip,keepaspectratio]{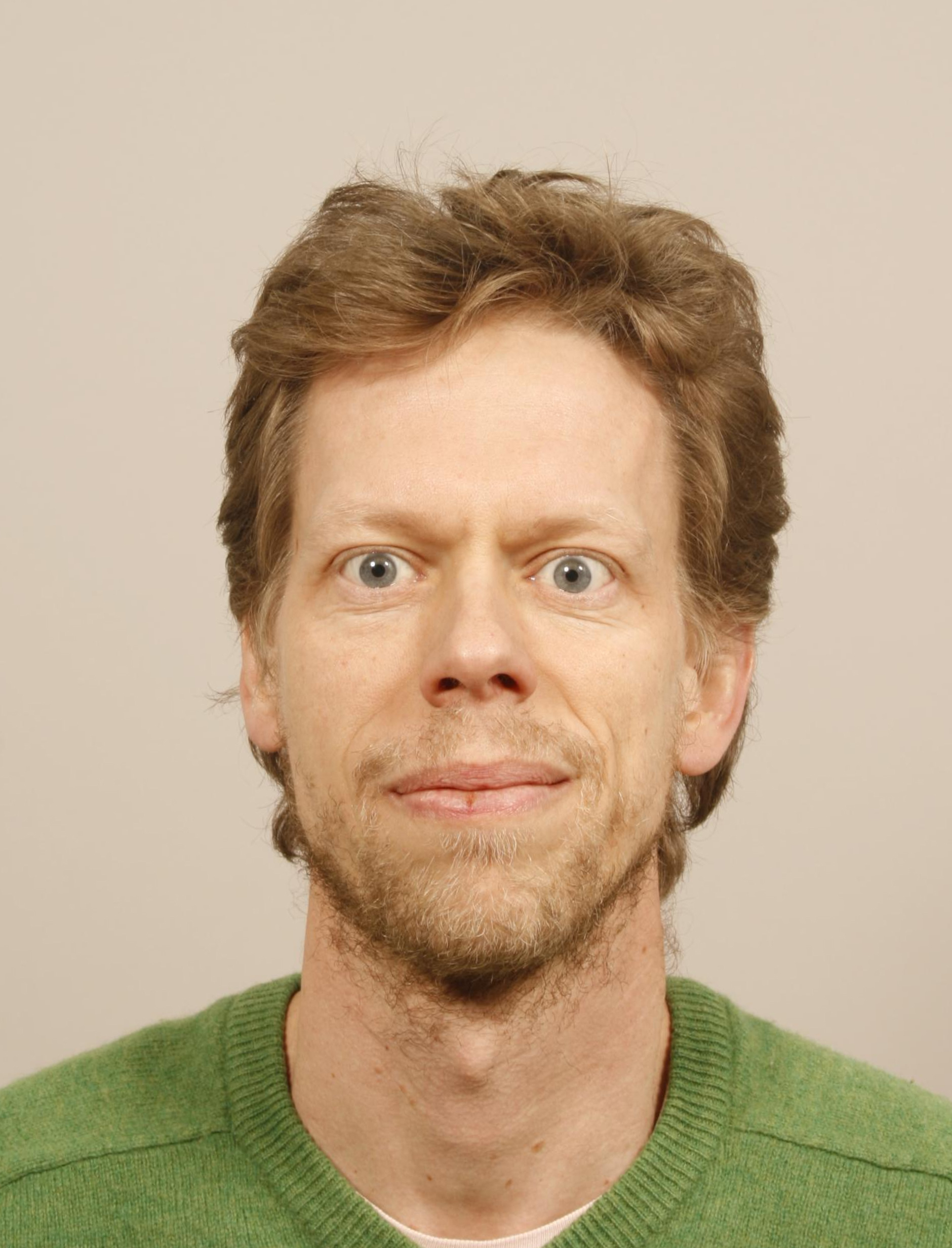}}]{Maurice H. ter Beek}
Maurice is a researcher at ISTI--CNR (Pisa, Italy) and head of the Formal Methods and Tools lab. He obtained a Ph.D.\ at Leiden University, The Netherlands. He has authored over 150~peer-reviewed papers, edited over 25~proceedings and special issues of journals, and he serves on the editorial board of the Journal of Logical and Algebraic Methods in Programming, Science of Computer Programming, PeerJ Computer Science and ERCIM News. He works on formal methods and model-checking tools for the specification and verification of safety-critical software systems and communication protocols, focusing in particular on applications in service-oriented computing, software product line engineering and railway systems. 
\end{IEEEbiography}




\end{document}